\documentclass[aps,showpacs,preprint,superscriptaddress,nofootinbib]{revtex4}

\begin{document}

\title{Dirac and  Lagrangian reductions in the canonical approach
to the first order form of the Einstein-Hilbert action }
\author{N. Kiriushcheva}
\email{nkiriush@uwo.ca}
\affiliation{Department of Mathematics,}
\author{S.V. Kuzmin}
\email{skuzmin@uwo.ca}
\affiliation{Department of Applied Mathematics, 
University of Western Ontario, London, N6A~5B7 Canada}
 
\date{\today}

\begin{abstract}
It is shown that the Lagrangian reduction, in which solutions of equations of 
motion that do not involve time derivatives are used to eliminate variables, 
leads to results quite different from the standard Dirac treatment of the 
first order form of the Einstein-Hilbert action when the equations of motion 
correspond to the first class constraints. A form of the first order 
formulation of the Einstein-Hilbert action which is more suitable for the 
Dirac approach to constrained systems is presented. The Dirac and reduced 
approaches are compared and contrasted. This general discussion is illustrated 
by a simple model in which all constraints and the gauge transformations which 
correspond to first class constraints are completely worked out using both 
methods in order to demonstrate explicitly their differences. These results 
show an inconsistency in the previous treatment of the first order 
Einstein-Hilbert action which is likely responsible for problems with its 
canonical quantization.


\end{abstract}

\maketitle

\section {Introduction}

Canonical quantization is the oldest, most rigorous, non-perturbative approach 
to quantization. It demands no new hypotheses (which is especially important 
in quantum gravity because of lack of experimental guides) and rests 
completely on the classical general theory of relativity and conventional 
methods of quantum field theory. For a discussion of the problems that one 
faces in trying to establish a connection between classical gravity and models 
built on new hypotheses, see the review \cite {Nicolai}. The first step in the 
canonical approach to any theory is to cast it into Hamiltonian form; analysis 
of this step is the main subject of this article.

The search for a canonical formulation of the Einstein-Hilbert (EH) action 
began after initial developments in analyzing the dynamics of singular (gauge 
invariant) systems where constraints arise 
\cite {Bergmann1951,Dirac1950,Dirac1951}.

Almost immediately after Dirac presented his work\footnote {The course of 
lectures given at Canadian Mathematical Seminar, Vancouver, August-September 
1949 and later published in \cite {Dirac1950,Dirac1951}.} on constraint 
dynamics the first attempt to apply his algorithm to the gravitational field 
was made by Pirani, Schild and Skinner \cite {PiraniSchild1950,
gamma-gamma}, and by Dirac himself \cite {Dirac1958,Dirac1959}.

In the above mentioned articles, the metric formulation of the EH action was 
used
\begin{equation}
\label{1}S_d(g^{\alpha \beta})=\int d^dx \sqrt{-g}g^{\mu \nu} R_{\mu \nu}
\left(\Gamma,\partial \Gamma\right),
\end{equation}
where $d$ is dimension of spacetime, $g=\det \left(g_{\alpha \beta}\right)$, 
affine connections $\Gamma_{\mu \nu}^\lambda$ are equal to Christoffel symbols 
$ \left\{_{\mu \nu }^\lambda \right\} = \frac{1}{2} g^{\lambda \sigma } 
\left( g_{\mu \sigma , \nu } + 
g_{\nu \sigma , \mu } - g_{\mu \nu , \sigma } \right)$ and $R_{\mu \nu}$ is 
the Ricci tensor expressed in terms of $\Gamma_{\mu \nu}^\lambda$ 
(see (\ref {3})). This is a ``second order'' formalism, as second derivatives 
of $g_{\mu \nu}$ appear in (\ref {1}).

Unlike `ordinary' gauge theories, the Dirac analysis \cite {Dirac1964} cannot 
be applied directly to (\ref {1}) because it is not known how to deal with 
second order derivatives using the Dirac procedure. (Both velocities {\it and} 
accelerations are explicitly present in (\ref {1}).) To avoid this problem, 
the so-called gamma-gamma form $L'_d$ \cite {LL} was used as a starting point 
in obtaining the Hamiltonian for pure gravity 
\begin{equation}
\label{2}L'_d\left( g^{\alpha \beta}\right)=
\sqrt {-g} g^{\alpha \beta }\left( \Gamma _{\sigma \lambda
}^\lambda \Gamma _{\alpha \beta }^\sigma -\Gamma _{\sigma \beta }^\lambda
\Gamma _{\alpha \lambda }^\sigma \right). 
\end{equation}

The Lagrangian of (\ref {2}) differs from that in (\ref {1}) by a total 
divergence \cite {LL}. The elimination of such a term does not affect the 
field 
equations but the reduced Lagrangian of (\ref {2}) is not relativistically 
invariant. (This was clearly stated in \cite {PiraniSchild1950} and reflected 
in its title: ``On the Quantization of Einstein's {\it Field Equations}'', not 
{\it action}.) The role of surface terms in the Hamiltonian formulation of 
General Relativity (GR) was discussed in \cite {RT1974}. Recently, the 
peculiar features of surface terms were reconsidered from quite different 
perspective in \cite {Padmanabhan2004} where it was demonstrated that it is 
not possible to obtain the full EH action (only its gamma-gamma part) starting 
from the standard graviton action built from non-interacting, massless, spin-2 
tensor field, and iterating in the coupling constant by having an interaction 
between the tensor field and its own energy-momentum tensor. 

The canonical approach based on (\ref {2}), instead of the EH action, is 
different from the canonical approach to ordinary gauge theories. If we use 
(\ref {2}), the invariance of original action is lost {\it completely}  and 
not just its {\it manifest} form as in ordinary gauge theories  after the time 
coordinate has been singled out. 

Dirac started his analysis using (\ref {2}) and later added the particular 
divergence term. According to \cite {ADM4}, this is a logically incomplete 
procedure. He also introduced space-like surfaces and fixation of coordinates 
in order to keep a space-like surface always space-like. This obviously 
destroys general covariance. In the conclusion to his paper \cite {Dirac1958} 
which is, probably, not well-known, Dirac clearly stated what one gives up  
in his formulation: ``One starts with ten degrees of freedom for each point in 
space, corresponding to the ten $g_{\mu \nu}$, but one finds with the method 
here followed that some drop out, leaving only six, corresponding to six 
$g_{rs}$. {\it This is a substantial simplification, but it can be obtained 
only at the expense of giving up four-dimensional symmetry.} I am inclined to 
believe from this that four-dimensional symmetry is not a fundamental property 
of the physical world.'' In the next paragraph he continued: ``The present 
paper shows that Hamiltonian methods, if expressed in their simplest form, 
{\it force one to abandon the four-dimensional symmetry}.'' (Italic of 
Dirac) This conclusion gives {\it only the relationship} between this 
simplest form of his Hamiltonian methods and four-dimensional symmetry. 
Accepting Dirac's conclusion means that GR has to be finally reformulated 
without four-dimensional symmetry. This is what is done in \cite {ABFKM2005}, 
where GR is reexpressed as a theory of evolving 3-dimensional conformal 
Riemannian geometries obtained by imposing two general principles: 1) time is 
derived from change; 2) motion and size are relative.

In contrast, if one believes that four-dimensional symmetry is a fundamental 
property of Nature and wants to keep this symmetry with the intention of 
eventually quantizing the EH action, one has to abandon the simplest 
Hamiltonian methods and try to find a Hamiltonian formulation that does not 
destroy four-dimensional symmetry right from outset. Two possible ways of 
doing this exist. The first one is to modify the Dirac procedure and work with 
the explicit dependence of the EH action on acceleration. The second is to 
find an {\it equivalent} formulation of the EH action that permits use of the 
standard Dirac procedure. (In addition to these orthodox approaches, 
there are a few more which are less developed; see p.54 of 
\cite {Thiemann2003LNP} and references therein.)

In the first case, we can consider the EH Lagrangian as a Lagrangian with 
higher derivatives and try to apply the Ostrogradsky Hamiltonian formulation 
\cite {Ostrogradskii} with appropriate adjustments to accomodate singular 
systems. (It was clearly indicated by Ostrogradsky that he considered only  
non-singular cases.) The first systematic generalization to singular cases was 
given by Gitman and Tyutin \cite {GitmanTyutin1983} (see also 
\cite {high-derivatives}). A full analysis of the EH action or some models  
where higher order derivatives enter {\it only} in such a way that they do not 
affect the equations of motion, to the best of our knowledge, does not exist. 
(The EH action in this respect is a kind of ``one and a half'' order system 
which probably creates problems in applying the Ostrogradsky method.) An 
attempt in this direction is due to Dutt and Dresden \cite {DuttDresden1986}.  

The second approach which does not involve reduction of the EH action by the 
elimination of a total divergence makes the action first order in 
derivatives by introducing auxiliary fields. If by elimination of these 
fields, we can return exactly to the original action (including terms with 
second order derivatives), we have an equivalent form. This form of the EH 
action is Einstein's affine-metric formulation \cite {EIN1925}; it is just 
linear in first order derivatives, so the standard Dirac procedure can be 
applied similarly to the way it is applied to a first order formulation of 
ordinary gauge theories. Moreover, all terms of the first order action 
contribute to the equations of motion, as opposed, to the second order 
formulation, and so the effect of all terms can be studied on the same footing.

Einstein considered $g^{\alpha \beta}$ and $\Gamma _{\alpha\beta }^\sigma$
as independent fields without assuming 
$\Gamma _{\alpha\beta }^\sigma=\left\{_{\alpha \beta }^\sigma \right\} $, 
since they are varied independently\footnote { This formulation was inspired by
his search for unification of gravity and electromagnetism (he originally 
tried to use the Eddington, pure affine, formulation \cite {EINEdd}), so, 
the symmetry of 
$g^{\alpha \beta}$ and $\Gamma _{\alpha \beta }^\sigma$ in $\alpha \beta$ 
was also lifted. (For further developments along this line see 
\cite {DDM1993}.) Einstein considered this formulation as the best starting 
point for possible generalizations of GR.}. 
In \cite{EIN1925}, he also proved that for symmetric 
$g^{\alpha \beta}$ and $\Gamma _{\alpha\beta }^\sigma$ this formulation is 
equivalent to (\ref {1}) and said that this was ``the most simple and 
consistent way'' of obtaining the field equations from the action principle. 
He also noted that with this formulation, no variation of fields on boundaries 
is needed (see also \cite {Wald}, Appendix E). In this approach the division 
of variables into being bulk or surface, as in \cite {RT1974}, 
is avoided and all variables are treated on the same footing with field 
variations vanishing on the boundary.

The Lagrange density of \cite {EIN1925} (eq.(3)) is given by 
\begin{equation}
\label{3}L_d(g^{\alpha \beta},\Gamma _{\alpha\beta }^\sigma)=
h^{\alpha \beta}R_{\alpha \beta}=
h^{\alpha \beta }\left( \Gamma _{\alpha \beta ,\lambda
}^\lambda -\Gamma _{\alpha \lambda ,\beta }^\lambda +\Gamma _{\sigma \lambda
}^\lambda \Gamma _{\alpha \beta }^\sigma -\Gamma _{\sigma \beta }^\lambda
\Gamma _{\alpha \lambda }^\sigma \right), 
\end{equation}
where $h^{\alpha \beta}=\sqrt{-g}g^{\alpha \beta }$ is just a simplifying 
notation and is not treated as an independent variable. (If we 
consider $h^{\alpha \beta}=\sqrt{-g}g^{\alpha \beta }$  as a change
of variables, the functional Jacobian $\left| \frac{\delta 
h^{\alpha \beta}}{\delta g^{\mu \nu}}\right|$ is field dependent in 
all dimensions $d>2$ and for $d=2$ is singular.) 
Moreover, if we consider $h^{\alpha \beta}$ as an independent field in 
(\ref {3}) without taking into account the field dependence of the Jacobian, 
we cannot return to $g^{\alpha \beta}$.  

The equivalence of (\ref {3}) to the second order form (\ref {1}) ($d>2$) 
follows from the solution of the field equations for 
$\Gamma_{\alpha \beta}^{\lambda}$, which is just 
the Christoffel symbol \cite {EIN1925}. We then obtain the standard Einstein 
field equations in terms of $g^{\alpha \beta}$. The first order Lagrangian 
reduces to the second order Lagrangian by substitution of the solution
$\Gamma_{\mu \nu}^\lambda =\left\{_{\mu \nu }^\lambda \right\}$ 
into the first order Lagrangian. 

A canonical analysis of the first order form of the EH action was given for 
the first time by Arnowitt, Deser and Misner (ADM) \cite{ADM1,ADM2}. (They 
also refer to some preliminary unpublished steps based on the first order 
action made by Schwinger.) However, in \cite{ADM1, ADM2} the Dirac procedure 
was not used and preliminary Lagrangian reduction was performed to 
obtain a {\it reduced Lagrangian} with fewer fields than are used in a 
canonical formulation. To do this reduction, the time independent equations 
of motion are solved to eliminate certain fields.  

Straightforward application of the Dirac procedure in the case of the first 
order formulation of gauge theories such as Maxwell theory \cite {Sund} is 
well-known. 
In this approach, conjugate momenta to all independent variables are 
introduced and this immediately produces an equivalent number of primary 
constraints as 
all velocities enter the Lagrangian only linearly. From this point, we follow 
the standard path by considering the conservation of constraints in time which 
produce secondary and higher constraints, until it is possible to have all 
constraints conserved. The $4D$ Maxwell Lagrangian gives 14 constraints, two 
of which are first class and the twelve remaining ones are second class 
\cite {Sund}. However, this is only a demonstration of the consistency of the 
Dirac procedure. The next step is the elimination of all second class 
constraints by passing from Poisson brackets (PB) to Dirac brackets.  
The Dirac reduction of Maxwell Lagrangian in first order form is performed in 
Appendix B providing a proof of the equivalence of using the second order and 
first order actions in a canonical analysis. 
 
A brief discussion of applying the Dirac approach to (\ref{3}) can be found 
in \cite{Sund} where expressions for the primary constraints are explicitly 
given, emphasizing that the first order formulation of the EH action in $4D$ 
results in 50 primary constraints, serving  as an illustration of the 
complexity of the Dirac procedure. The number of independent field components 
of $g^{\alpha \beta}$, $\Gamma_{\alpha \beta}^{\lambda}$ in (\ref {3}) is 
$\frac {1}{2}d\left(d+1\right)^2$ in $d$ dimensions, and introducing 
conjugate momenta doubles the number of phase-space variables. This large 
number is a way of showing the complexity of the EH action, but this is not a 
real  problem, as in the Hamiltonian analysis we separate only spatial and 
temporal indices of fields so that, in this case, we have only nine distinct 
fields for all $d$: $g^{00},g^{0k},
g^{km},\Gamma_{00}^{0},\Gamma_{0k}^{0},\Gamma_{km}^{0},\Gamma_{00}^{k},
\Gamma_{0m}^{k},\Gamma_{mn}^{k}$. This is not greatly different from using 
four fields in the first order formulation of electrodynamics $A_{0},A_{k},
F_{0k},F_{km}$. 

In the $2D$ limit, the first order action is not equivalent to the second 
order action, which is a total divergence \cite {LL}. This was analyzed at the 
level  of the Lagrangian in \cite {Mann}. The first order Lagrangian in $2D$ 
is not a total divergence and its canonical form can be discussed just like 
any other model of $2D$ gravity \cite {2Dmodels}. Moreover, the first order 
formulation as a general field-theoretical construction should be valid in all 
dimensions, with possibly special behaviour in some particular dimensions, 
but also with some similarities in all dimensions\footnote {The possibility of 
similarities of the $2D$ limit of the first order form of the EH action with 
the higher dimensional form has to be stronger than is possible in the case of 
electrodynamics. In the $2D$ limit of the first order form of the EH action we 
have nine distinct types of fields just as in all higher dimensions while in 
electodynamics there are only three fields in $2D$ as opposed to four in 
higher dimensions (as in $2D$, $F_{km}=0$).}. 

These considerations have motivated us to perform a canonical analysis of 
the $2D$ EH action using the Dirac procedure without any {\it a priori} 
assumptions or restrictions. In particular, it is important to find the 
algebra of constraints. In Dirac's analysis of GR \cite {Dirac1958} the PB 
algebra of constraints \cite {Dirac1964} is non-local with field dependent 
structure constants
\begin{equation}
\label{Dirac1}\left\{H_{a}\left(x\right),H_{b}\left(x'\right)\right\}
=H_{b}\left(x\right)\delta_{,a}\left(x,x'\right)-\left(ax \leftrightarrow bx' 
\right),
\end{equation}
\begin{equation}
\label{Dirac2}\left\{H_{a}\left(x\right),H\left(x'\right)\right\}
=H\left(x \right)\delta_{,a}\left(x,x'\right),
\end{equation}
\begin{equation}
\label{Dirac3}\left\{H\left(x\right),H\left(x'\right)\right\}
=h^{ab}\left(x \right)H_{a}\left(x\right)\delta_{,b}\left(x,x'\right)
-\left(x \leftrightarrow x'\right).
\end{equation}

This type of algebra (sometimes called {\it a hypersurface deformation 
algebra} \cite {Thiemann2001}) is not encounted in `ordinary' gauge theories. 
This is not a true Lie algebra, this being the main obstacle to canonically 
quantizing GR. The 
question that arises is whether this is an intrinsic property of GR or the 
result of assumptions made in the course of analyzing the reduced Lagrangian 
in the approach of Dirac \cite {Dirac1958,Dirac1959} and ADM \cite {ADM1,ADM2}.
It turns out that in the $2D$ case, the Dirac procedure gives a local algebra 
of constraints with field independent structure constants 
\cite{KKM1}. In order to preserve `ordinary' properties beyond 
locality of the PB and to have also off-shell closure of the PB algebra of 
generators 
and off-shell invariance of the Lagrangian, we have made a simple linear 
transformation of the affine connections.  
In \cite {KKM1}, such transformations were expressed in component form but 
in fact they can be recast in the covariant form
\begin{equation}
\label{4}\xi_{\alpha \beta}^\lambda = \Gamma _{\alpha \beta}^\lambda 
-\frac{1}{2}\left(\delta_\alpha^\lambda \Gamma _{\beta \sigma}^\sigma + 
\delta_\beta^\lambda \Gamma_{\alpha \sigma}^\sigma \right). 
\end{equation}

This covariant change of variables is quite different from the usual 
non-covariant change and it provides an alternative covariant formulation
of the first order form of the EH action which is more suitable for canonical 
analysis than the form of (\ref {3}). We have not been able to find any 
particular geometrical significance of the variables 
$\xi_{\alpha \beta}^\lambda$
but it appears that they reflect the dynamical properties of {\it fields} of 
the first order EH action which is richer than the geometrodynamics of 
space-like surfaces. According to Hawking, using a family of space-like 
surfaces is in contradiction to the whole spirit of General Relativity and 
restricts the topology of spacetime \cite {Hawking}. (This echoes Dirac's 
conclusion in \cite {Dirac1958}, partially cited above.) This restriction, 
imposed by the slicing of spacetime, must be lifted at the quantum level 
\cite {Thiemann2001}; avoiding it at the outset seems to be the most 
natural cure of this problem. The idea of slicing spacetime originated in the 
attempt ``to recover the old comforts of a Hamiltonian-like scheme: a system 
of hypersurfaces stacked in a well defined way in spacetime, with the system 
of dynamical variables distributed over these hypersurfaces and developing 
uniquely from one hypersurface to another'' \cite {Kuchar}. This, although 
`reasonable' from the point of view of classical Laplacian determinism, is 
hard to justify from the standpoint of General Relativity 
\cite {HawkingPenrose1970}. In GR, an entire spatial slice can only be seen by 
an observer in the infinite future \cite {Fotini1998} and 
an observer at any point of a space-like surface cannot have information about 
the rest of a surface. (This actually follows from just the basic 
principles of relativity, those of locality and the finite speed of signals 
(e.g. see p.7 of \cite {LL}).) 
It would be unphysical to build any formalism by basing it on the 
development in time of data that can be available only in the infinite future 
and to try to fit GR into a scheme of classical determinism and 
non-relativistic Quantum Mechanics with its notion of a wave-function defined 
on a space-like slice. This idea also contradicts the canonical treatment of 
local relativistic field theories which do not make any explicit references to 
the ambient space-time by making use of a particular coordinate system or 
class of coordinate systems \footnote {The condition that a space-like surface 
remains space-like obviously imposes restriction on possible coordinate 
transformations, thereby destroying four-symmetry.}.

The change of variables of (\ref {4}) can be used in any dimension and 
it is quite natural to explore this change in higher dimensions with hope 
that, as in the $2D$ case, it leads to an algebra of constraints that has the 
form of a Lie algebra or to see how the non-locality associated with the 
``hypersurface deformation algebra'' appears in higher dimensions without 
imposing it from outset by choosing a particular slicing of spacetime.

In the next section we consider the effect of using 
$\xi_{\alpha \beta}^{\lambda}$ in place of $\Gamma_{\alpha \beta}^{\lambda}$ 
in any dimension and demonstrate that straightforward application of the Dirac 
procedure is considerably simplified by this choice of variables. After a few 
simple steps using Dirac reduction to eliminate second class constraints, 
we face sharp discrepancies with previous results \cite {ADM2} obtained by 
using Lagrangian reduction in which time independent equations of motion 
are used to eliminate some variables. The source of this difference and the 
conditions under which the two approaches are equivalent are analyzed. The 
next two sections provide the full canonical analysis of a simple model both  
using the Dirac approach (Sec.3) and using the Lagrangian reduction (Sec.4) 
in a way similar to \cite {ADM2} in order to illustrate the general 
considerations of Sec.2. The results are summarized in a conclusion. In 
Appendix A an alternative first order formulation of the EH action in which 
the variables $\xi_{\alpha \beta}^{\lambda}$ are used is demonstrated to be 
equivalent to the second order form of the EH action. In Appendix B we perform 
the Hamiltonian (Dirac) reduction with the first order formulation of Maxwell 
electrodynamics by eliminating those secondary constraints that are of a 
special form (this is an illustration of what was done in the  EH action in 
Sec.2 and in a simple model in Secs.3 and 4) and prove in this way that, 
starting from the first order form, one can obtain all of the standard results 
usually derived using the second order form of the action. In Appendix C,  
Lagrangian reduction of the first order form 
of the EH action in any dimension based on the variables 
$\xi_{\alpha \beta}^{\lambda}$ is performed in a way consistent with the Dirac 
analysis.

\section {Canonical analysis of first order form of the EH action in any 
dimension}

In this section we discuss the Hamiltonian formulation of the EH action using 
the generalization of the transformation of (\ref {4}) that produces  
canonical results similar to those of ordinary gauge theories in the $2D$ 
limit of the first order form of the EH action. The inverse transformation of 
(\ref {4}) in $d$ dimensions is given by
\begin{equation}
\label{1.4}\Gamma_{\alpha \beta}^\lambda = \xi _{\alpha \beta}^\lambda 
-\frac{1}{d-1}\left(\delta_\alpha^\lambda \xi _{\beta \sigma}^\sigma + 
\delta_\beta^\lambda \xi_{\alpha \sigma}^\sigma \right)
\end{equation}
which upon substitution into (\ref {3}) gives 
\begin{equation}
\label{1.5}\tilde L_d\left(g,\xi\right) = h^{\alpha \beta}\left( 
\xi _{\alpha \beta,\lambda }^\lambda -
\xi _{\alpha \sigma}^\lambda \xi_{\beta \lambda }^\sigma + 
\frac{1}{d-1} \xi _{\alpha \lambda }^\lambda \xi_{\beta \sigma }^\sigma
\right),  
\end{equation}
an alternative first order form of the EH action. This is because the 
linear transformation used for the field $\Gamma_{\alpha \beta}^\lambda$ 
appears in (\ref {3}) at most bilinearly and only linearly in their 
derivatives. It is also possible to prove the equivalence of the first order 
form (\ref {1.5}) with second order form (\ref {1}) by solving the equation of 
motion for $\xi_{\alpha \beta}^\lambda$, 
and substituting the resulting expression for $\xi_{\alpha \beta}^\lambda$ 
into the equation of motion for $g^{\mu \nu}$ \footnote {This is similar to 
what was done to prove the equivalence of (\ref {3}) to (\ref {1}) in 
\cite {EIN1925}.}. As a result, we obtain the Einstein field equations without 
any reference to the affine connection. Actually, solving the equation of 
motion for $\xi_{\alpha \beta}^\lambda$ is simpler than solving that of 
$\Gamma_{\alpha \beta}^\lambda$. (Details are given in Appendix A.)

However, the main advantage of (\ref {1.5}) is that it is
extremely well suited for applying the canonical procedure. There is now  nice 
separation of components of $\xi_{\alpha \beta}^\lambda$ into those which are 
dynamical and non-dynamical, as the only term with derivatives is of the form
\begin{equation}
\label{1.6}\xi _{\alpha \beta,\lambda }^\lambda =
\dot \xi _{\alpha \beta}^0 + \xi _{\alpha \beta, k }^k.  
\end{equation}
(Latin indices are spatial and a dot represents a time derivative.) Following 
Dirac, the first step is to introduce momenta conjugate to all fields
\begin{equation}
\label{1.Md}\pi_{\alpha \beta}\left( g^{\alpha \beta}\right), 
\Pi_{0}^{\alpha \beta}\left( \xi_{\alpha \beta}^{0}\right), 
\Pi_{k}^{\alpha \beta}\left( \xi_{\alpha \beta}^{k}\right).
\end{equation}

Using (\ref {1.5}), we immediately obtain the primary constraints 
\begin{equation}
\label{1.7}\pi _{\alpha \beta} \approx 0,\Pi_k^{\alpha \beta} \approx 0,
\Pi_0^{\alpha \beta} - \sqrt{-g} g^{\alpha \beta} \approx 0
\end{equation}
which equals the number of fields in the Lagrangian.

The total Hamiltonian is
$$
H_T=H_c+\lambda^{\alpha \beta}\pi_{\alpha \beta} + 
\Lambda_{\alpha \beta}^0 \left(\Pi_0^{\alpha \beta}-h^{\alpha \beta}\right)+
\Lambda_{\alpha \beta}^k \Pi_k^{\alpha \beta},
$$
\begin{equation}
\label{1.8}H_c=-h^{\alpha \beta}\left( 
\xi _{\alpha \beta, k }^k -
\xi _{\alpha \sigma}^\lambda \xi_{\beta \lambda }^\sigma + 
\frac{1}{d-1} \xi _{\alpha \lambda }^\lambda \xi_{\beta \sigma }^\sigma
\right),   
\end{equation}
where $\lambda^{\alpha \beta}$ and $\Lambda_{\alpha \beta}^{\gamma}$ are 
Lagrange multipliers associated with the primary constraints.

If the $d(d+1)$ by $d(d+1)$ matrix 
\begin{equation}
\label{1.10}\tilde M_d=\left( \left\{\phi,\tilde \phi \right\} \right) 
\end{equation}
built from the non-zero PB among the primary constraints 
$\left( \phi,\tilde \phi \in \left(\pi _{\alpha \beta}, \Pi_0^{\gamma \sigma} 
- \sqrt{-g} g^{\gamma \sigma}\right) \right) $ is invertible, we have a subset 
of constraints which are second class. Moreover, all these constraints are of 
a special form involving the canonical pair 
$(g^{\alpha \beta },\pi_{\alpha \beta })$ for which 
$\pi_{\alpha \beta}\approx 0$ (see Dirac \cite{Dirac1964}, Appendix B, and for 
more detailed and general discussion \cite{GT1990}). For such constraints, if 
$\det \tilde M_d\neq 0$, we can set all momenta $\pi_{\alpha \beta }$ to 
zero and then solve $ \Pi_0^{\gamma \sigma} = \sqrt{-g} g^{\gamma \sigma}$ 
for $g^{\alpha \beta }=g^{\alpha \beta }\left( \Pi_0^{\gamma \sigma}\right)$
and use this equality to eliminate $g^{\alpha \beta}$ in both the Hamiltonian 
and the remaining constraints. This is Hamiltonian (Dirac) reduction in its  
simplest form. Dirac brackets are equal to PB for all the remaining variables. 
The use of this reduction is shown in Appendix B to lead to equivalence of the 
first and second order formulations for electrodynamics at the level of the 
Hamiltonian.

Actually, for $\tilde L_d$ it is not even necessary to solve the equations 
$\Pi_0^{\gamma \sigma} = \sqrt{-g} g^{\gamma \sigma}$ for 
$g^{\alpha \beta }$ as they enter the Hamiltonian in the particular 
combinations which are present in the second class  primary constraints and 
the solution for such combinations are, of course, obvious if the 
condition $\det \tilde M_d\neq 0$ is fulfilled. In this case, the canonical 
analysis of the reduced Hamiltonian leads to the form of the gauge 
transformation of $\Pi_0^{\alpha \beta}$ and so using the strong equality, 
$\Pi_0^{\alpha \beta}=\sqrt{-g}g^{\alpha \beta}$, we can immediately find the 
gauge transformation of $g^{\alpha \beta}$.

In $2D$ (and only in $2D$) the matrix (\ref {1.10}) is singular. The rank of 
$\tilde M_2$ is four and its dimension is $6$ by $6$, so only two pairs of 
constraints constitute a subset of second class constraints which are of a 
special form meaning they can be eliminated. (For more details see 
\cite {KKM2}.)

Let us denote the number of independent components of a field $\Phi$ by 
$[\Phi]$. At this stage, a reduction of the 
$[g^{\alpha \beta}]=\frac{1}{2}d(d+1)$ fields has been performed by 
eliminating the canonical pairs ($g^{\alpha \beta},\pi_{\alpha \beta}$) 
using the primary second class constraints. The remaining primary constraints 
$\Pi_k^{\alpha \beta}$ will produce secondary constraints 
($\chi_k^{\alpha \beta}$)
\begin{equation}
\label{1.11}\dot \Pi_k^{\alpha \beta}=\left\{ \Pi_k^{\alpha \beta},H_c
\right\}=-\frac{\delta H_c}{\delta \xi_{\alpha \beta}^k}\equiv
\frac{\delta \tilde L_d}{\delta \xi_{\alpha \beta}^k}=\chi_k^{\alpha \beta}  
\end{equation}
with $[\chi_k^{\alpha \beta}]=\frac{1}{2}(d-1)d(d+1)$.

Explicitly separating time and space indices, we obtain three secondary 
constraints ($\chi_k^{mn}, \chi_k^{0m}, \chi_k^{00}$)
\begin{equation}
\label{1.12}\frac{\delta \tilde L_d}{\delta \xi_{mn}^k}=\chi_k^{mn}=
-h_{,k}^{mn}-h^{\mu m}\xi_{\mu k}^{n}-h^{\mu n}\xi_{\mu k}^{m}+\frac {1}{d-1}
\left(h^{\mu m}\xi_{\mu \lambda}^{\lambda}\delta_{k}^{n} 
+h^{\mu n}\xi_{\mu \lambda}^{\lambda}\delta_{k}^{m}\right),
\end{equation}
\begin{equation}
\label{1.13}\frac{\delta \tilde L_d}{\delta \xi_{0m}^k}=\chi_k^{0m}=
-h_{,k}^{0m}-h^{\mu 0}\xi_{\mu k}^{m}-h^{\mu m}\xi_{\mu k}^{0}+\frac {1}{d-1}
h^{\mu 0}\xi_{\mu \lambda}^{\lambda}\delta_{k}^{m}, 
\end{equation}
\begin{equation}
\label{1.14}\frac{\delta \tilde L_d}{\delta \xi_{00}^k}=\chi_k^{00}=
-h_{,k}^{00}-2 h^{00} \xi_{0k}^0-2 h^{0m} \xi_{mk}^0. 
\end{equation}

From the point of view of the Dirac procedure, these three constraints are 
quite different. The distinction that arises between these constraints is not 
taken into account when they are treated as time independent Lagrangian 
equations of motion in the ADM approach to the first order Lagrangian of 
(\ref {3}) \cite {ADM2,Fad1982}. The matrix of PB of $\chi_k^{mn}$ with 
the corresponding primary constraints $\Pi_k^{mn}$ is non-singular. These 
constraints form a second class subset and this is a subset of the same 
special form as the part of second class primary constraints that have been 
already considered. Following Dirac reduction, we have $\Pi_k^{mn} =0$ and 
$\xi_{mn}^k = \xi_{mn}^k(\xi_{\alpha \beta}^0,\xi_{0p}^q)$ - solutions
of the second class constraints $\chi_k^{mn}=0$ that are now substituted into 
the Hamiltonian and remaining constraints. For the second equation 
(\ref {1.13}), the matrix of PB of $\chi_k^{0m}$ with the correspondent 
primary constraints ($\Pi_k^{0m}$) is singular and this subset is not purely 
second class. According to Dirac, we have to find the maximum possible number 
of first class combinations for this subset and only the remaining 
constraints which are second class can be eliminated. Among the constraints 
$\chi_k^{0m}$ of (\ref {1.13}), only one first class combination exists and it 
is $\chi_k^{0k}$ 
\begin{equation}
\label{1.13a}\chi_k^{0k}=
-h_{,k}^{0k}-h^{mk}\xi_{mk}^{0}+h^{00}\xi_{00}^{0}. 
\end{equation}

Only the fields $\xi_{\alpha \beta}^{0}$ are present in (\ref {1.13a}) and 
they have a vanishing PB with the primary constraints $\Pi_k^{\alpha \beta}$.
The remaining constraints are again of the special form, so we can further 
reduce our system by eliminating $[\chi_k^{0m}]-1=(d-1)^2-1$ fields.  
Elimination of these fields without destroying tensorial notation is performed 
in Appendix C.

The last constraint (\ref{1.14}) is first class as there are no components of 
$\xi_{\alpha \beta}^k $ appearing in (\ref{1.14}), but only these components 
give 
a non-zero PB with the primary constraints $\Pi_k^{\alpha \beta} $. Dirac 
reduction using (\ref {1.12}-\ref {1.14}) leads to the following number of 
fields in the reduced Hamiltonian
\begin{equation}
\label{1.15}[g^{\alpha \beta}]+[\xi_{\alpha \beta}^{\gamma}]-[\chi_k^{mn}]-
[\chi_k^{0m}]+1=d(d+2). 
\end{equation}

Taking into account this reduction using the second class subset of primary 
constraints, we have only $d$ primary constraints ($[\Pi_k^{00}]+1$) and $d$ 
secondary constraints left.

The secondary constraints are now
$$
\chi_k^{00}=-\Pi_{0,k}^{00}-2 \Pi_0^{00} \xi_{0k}^0-2 \Pi_0^{0m} \xi_{mk}^0,
$$ 
\begin{equation}
\label{1.16}\chi_{k}^{0k}=-\Pi_{0,k}^{0k}-\Pi_0^{nm} \xi_{nm}^0+ \Pi_0^{00} 
\xi_{00}^0  
\end{equation}
where we have used the strong equality $h^{\alpha \beta}=
\Pi_0^{\alpha \beta}$. These 
are a $d$ dimensional generalization of two of the three constraints found 
in $2D$ \cite {KKM1}, having a simple local PB algebra 
\begin{equation}
\label{1.16a}\left\{ \chi_{k}^{0k} \left( x \right) , 
\chi_n^{00} \left( y \right) \right\} =\chi_n^{00}\left( x \right)\delta^{d-1} 
\left( x-y \right), 
\left\{ \chi_{k}^{0k}, \chi_k^{0k}\right\}=0,
\left\{ \chi_{k}^{00},\chi_n^{00}\right\}=0
\end{equation}
and zero PB with primary constraints. At this stage the primary and secondary 
constraints form a first class system. We now continue the Dirac procedure; it 
leads to the existence of, at least, tertiary constraints.

Already after the first steps of Dirac reduction using 
$\xi_{\alpha \beta}^{\gamma}$ and $g^{\alpha \beta}$ as 
independent variables, we see that primary and secondary constraints having 
a local PB algebra with field independent structure constants arise and that 
tertiary constraints must be present in the Hamiltonian, which is no longer a 
linear combination of secondary constraints as in the $2D$ case 
\cite {KKM1,KKM2}. 
This result is quite unlike the previous treatment of the first order EH 
Lagrangian \cite {ADM2} where after Lagrangian reduction the Hamiltonian is a 
linear combination of secondary constraints with a non-local hypersurface 
deformation algebra of constraints with field dependent structure constants. 
This has been viewed as an inconsistency in the constraint algebra and the 
main obstacle to canonically quantizing GR \cite{Fotini1996, Thiemann2003} or 
as an indication of the non-locality of Nature and an inspiration for new 
ideas, such 
as promoting this algebra to being a first principle, more fundamental than 
the action principle or the equations of motion \cite{newideas}.

Recently, Kummer and Sch\"utz have reconsidered the first order formulation of 
$4D$ GR using Cartan variables \cite {Kummer}. Their analysis is based on 
avoiding the ADM decomposition that has been almost exclusively used when 
discussing tetrad gravity. Their approach also leads to tertiary constraints 
and a local algebra of constraints. 

Before continuing with the Dirac procedure it is necessary to understand why 
the two approaches, Dirac and Lagrangian reduction, that are supposed to be 
equivalent lead to different results. We attempt to answer this question in 
the rest of this paper. 
 
First of all, let us note that the presence of tertiary constraints is not in 
contradiction with the number of degrees of freedom. For example, if tertiary
constraints are all first class and the Dirac procedure is closed at this 
stage, we have $3d$ first class constraints. The number of fields in the 
reduced Hamiltonian is $d(d+2)$ (see (\ref {1.15})) minus $\frac{1}{2}d(d+1)$ 
because of the first reduction using the second class primary constraints. 
The result is $\frac{1}{2}d(d-3)$, the number of degrees of freedom associated 
with a symmetric tensor gauge field in $d$ dimensions. This expression works 
only for $d>2$; $2D$ is a special case which cannot be described by this 
relation because there are no second class constraints among the secondary 
constraints. (See \cite {KKM2} for a full discussion of the $2D$ EH action 
with $g^{\alpha \beta}$ being treated as an independent field.)
Of course, having $3d$ first class constraints is not the only possibility 
that results in the expected number of degrees of freedom, but this 
demonstrates that the presence of tertiary constraints is not inconsistent.
Moreover, it is necessary to have tertiary constraints, because without 
solving the first class constraints we have additional independent variables 
and extra constraints are needed to reduce the number of degrees of freedom to
the expected value.  

Secondly, Lagrangian and not Dirac reduction was used in \cite{ADM2}. (For 
a very clear exposition of this reduction see Appendix A of the review article 
\cite {Fad1982}.) 

In the Dirac approach, after identifying all second class constraints of the 
special form among the constraints (\ref {1.12}-\ref {1.14}) and then 
eliminating the corresponding variables, we reduce the number of fields (see 
(\ref {1.15})) to just $d(d+2)$ which is $24$ if $d=4$. By way of contrast, in 
\cite {ADM2} the solution of the 30 Lagrangian equations of motion that do not 
involve time derivatives of any component of the affine connection leads to a
reduced  Lagrangian with only 16 independent variables (see eq.(4.1) of 
\cite{ADM2} and also eq.(A.27) of \cite {Fad1982}), so that when using 
Lagrangian reduction 34 variables have disappeared after solving only 30 
equations of motion. This is clear indication that secondary first class 
constraints have been solved in this approach. When using the variables 
$\xi_{\alpha \beta}^{\gamma}$, if we were to use the solutions of the first 
class constraints of (\ref {1.16}) to eliminate fields, we eliminate more 
variables from the Lagrangian than equations 
of motion that have been solved. This contradicts the Dirac prescription for 
treating 
constrained dynamical systems and illustrates the importance of classifying 
constraints into first and second class, though the importance of this 
classification has been deemphasized in \cite{FJ}.

The Lagrangian reduction of the Maxwell action written in first order form is, 
by way of contrast, fully justified as all time independent equations of 
motion in this case correspond to second class constraints 
\footnote {The usual references (e.g., \cite {ADM1}) concerning the similarity 
of this reduction to reduction of the first order form of the EH action 
are not entirely correct.}. The equations of motion which do not have time 
derivatives (the Lagrangian constraints) for the first order formulation of 
electrodynamics are
\begin{equation}
\label{1.17}\frac{\delta L_M}{\delta F^{km}}=F_{km}-\left(\partial_k A_m-
\partial_m A_k\right) 
\end{equation}
giving $F_{km}$ immediately in terms of $A_m$. This is in agreement with Dirac 
reduction, as the primary constraints $\Pi^{km}\approx 0$ (where $\Pi^{km}$ 
are the momenta conjugate to $F_{km}$) give a subset of the second class 
constraints of the special form which allows for Dirac elimination or ensures 
that the reduced Lagrangian is equivalent to the initial one (see Appendix B). 

Full correspondence between two reduction procedures exists only if we 
eliminate variables using Lagrangian equations similar to (\ref {1.17}) where 
the field being eliminated is the same as the field being varied. (This also 
is the situation for auxiliary fields in supersymmetric models.) Only in this 
case is the reduced Lagrangian equivalent to the original. We see therefore 
that elimination of variables in the original Lagrangian by merely solving the 
Lagrange constraints may not always be correct. For example, suppose we have 
an action 
functional $S(Q,q)$ and that the equation $\delta S/\delta Q=0$ can be solved 
for the $Q$'s so that $Q=Q(q)$. This is then substituted back into $S$ and the 
new action $S'(q)=S(q,Q(q))$ implies a dynamical equation 
$\delta S'/\delta q=0$ which is the same dynamical equation for $q$ that 
follows from $\delta S/\delta q=0$. However, 
if $\delta S/\delta Q=0$ is solved for $q$ instead of $Q$ one does not, in 
general, obtain the same dynamical equations from the action obtained by 
substitution of this solution into the original action. (We illustrate this in 
Section 4.)

When using the variables  $\Gamma$, it is difficult to compare results of 
these two approaches since the straightforward Dirac procedure is not easy to 
apply because of the way the Lagrangian (\ref {2}) depends on derivatives of 
$\Gamma$. The subset of second class primary constraints 
used in \cite {KKM1}, where the variables $\Gamma$ are employed, leads to 
elimination of some of the variables, which  affect the rest of the primary 
constraints and some primary constraints become combinations of momenta and 
not just simply $\Pi_k^{\alpha \beta}$, as in the case when using the 
variables 
$\xi$. Consequently not all secondary constraints have the same simple 
special form structure with primary constraints. In Lagrangian reduction, 
solving 
those equations of motion without time derivatives for auxiliary fields such 
as is done in (\ref {1.17}), is not as easy to analyze because the last 
equality 
in (\ref {1.11}) is due to the ``diagonal'' form of terms with derivatives,  
which is not the case when one uses $\Gamma$ instead of $\xi$. 
 
To illustrate this general discussion we do not need to consider the EH action 
in $d$ dimensions (using either $\Gamma $ or $\xi $); we just need a simple 
model in which the first class constraints are present and can be 
algebraically solved in order to compare the results of the two approaches.
We consider $\tilde L_2\left(h, \xi\right)$ which is a simple model with first 
class constraints that can be algebraically solved \footnote {The 
non-equivalence of this model to the second order EH action in $2D$, 
the difference between the $h$ and $g$ formulations in $2D$, etc. are 
irrelevant since we are only comparing different {\it methods} of reduction.}. 
It has stronger connection 
with $\tilde L_d\left(g, \xi\right)$ for $d>2$ than 
$\tilde L_2\left(g,\xi\right)$. (To see this, compare the analysis of 
$\tilde L_2(h, \xi)$ in \cite {KKM1} with that of $\tilde L_2(g, \xi)$ in 
\cite {KKM2}).

In the next two sections we present the complete canonical analysis of this 
simplest Lagrangian using both approaches. We examine the constraint structure 
and analyze the invariance of the action under the gauge transformation that 
is implied by the full set of first class constraints using the approach of 
Castellani \cite {Castellani}. 

\section {Canonical analysis of a simple model using Dirac reduction}

In this section, we give the full Dirac analysis of a slightly modified form 
of (\ref {1.5}) which has been advocated by Faddeev \cite {Fad1982}. It allows 
us to demonstrate the effect of neglecting the contributions of surface terms. 
We consider
\begin{equation}
\label{18} \tilde L_d^{'}=\tilde L_d-\left(h^{\alpha \beta} 
\xi _{\alpha \beta}^\lambda\right)_{,\lambda} = -h_{,\lambda}^{\alpha \beta} 
\xi _{\alpha \beta }^\lambda -
h^{\alpha \beta} \xi _{\alpha \sigma}^\lambda \xi_{\beta \lambda }^\sigma + 
\frac{1}{d-1} h^{\alpha \beta} \xi _{\alpha \lambda }^\lambda \xi_{\beta 
\sigma }^\sigma
\end{equation}
that in the $2D$ case results in 
\begin{equation}
\label{19}\tilde L_2^{'}=-\dot h^{11} \xi_{11}^0 - 2 \dot h^{01} 
\xi _{01}^0 - \dot h^{00} \xi_{00}^0 - H_c
\end{equation}
where
$$
H_c = \xi_{11}^1\left( h_{,1}^{11} - 2 h^{11}\xi_{01}^0 - 2 h^{01} 
\xi_{00}^0 \right)
$$
\begin{equation}
\label{20}+2\xi_{01}^1\left( h_{,1}^{01} + h^{11} \xi_{11}^0 - 
h^{00} \xi_{00}^0\right) +\xi _{00}^1 \left( h_{,1}^{00} + 2 h^{01} 
\xi_{11}^0 + 2 h^{00}\xi_{01}^0 \right). 
\end{equation}

Introducing conjugate momenta to all variables, $\pi_{\alpha \beta} 
\left(h^{\alpha \beta}\right)$ and $\Pi_\gamma^{\alpha \beta}\left(
\xi_{\alpha \beta}^\gamma\right)$, we immediately obtain the primary 
constraints 
\begin{equation}
\label{21}\Phi_{\alpha \beta} = \pi_{\alpha \beta}+\xi_{\alpha \beta}^0
\approx 0,\Pi_\gamma^{\alpha \beta} \approx 0  
\end{equation}
and the total Hamiltonian
\begin{equation}
\label{22}H_T=H_c+\lambda^{\alpha \beta}\Phi_{\alpha \beta} + 
\Lambda_{\alpha \beta}^\gamma \Pi_\gamma^{\alpha \beta}.  
\end{equation}

Among the primary constraints, we have a subset which is second class as
\begin{equation}
\label{23}
\left\{ \Phi_{\alpha \beta } ,\Pi _0^{\mu \nu }\right\} = \Delta ^{\mu \nu}_
{\alpha \beta }. 
\end{equation}

We are using the standard fundamental PB for independent fields
$$\left\{ h^{\alpha \beta },\pi _{\mu \nu }\right\} = \Delta _{\mu \nu
}^{\alpha \beta }, \left\{ \xi_{\alpha \beta }^\lambda ,\Pi _\sigma^
{\mu \nu }\right\} = \delta_\sigma^\lambda \Delta ^{\mu \nu}_{\alpha \beta },$$
where $\Delta _{\mu \nu}^{\alpha \beta} = \frac{1}{2} \left(\delta_\mu^\alpha 
\delta_\nu^\beta + \delta_\mu^\beta \delta_\nu^\alpha \right)$. The 
constraints (\ref {23}) are of a special form and, according  
Dirac reduction, such constraints can be eliminated without affecting the PB 
of all the remaining variables. We have now two strong equalities
\begin{equation}
\label{24}\Pi_0^{\alpha \beta}=0,\xi_{\alpha \beta}^0 =-\pi_{\alpha \beta} 
\end{equation}
and as a result, we have the reduced total Hamiltonian
$$
H_T^{(1)}=H_c^{(1)} + \Lambda_{\alpha \beta}^k \Pi_k^{\alpha \beta},
$$
\begin{equation}
\label{25}
H_c^{(1)}=-\xi_{11}^1 \tilde \chi_1^{11}-2\xi_{01}^1 \tilde \chi_1^{01}-
\xi_{00}^1 \tilde \chi_1^{00}
\end{equation}
where using (\ref {24})
$$
\tilde \chi_1^{11}=-\left(h_{,1}^{11}+2 h^{11} \pi_{01}+2 h^{01} \pi_{00}
\right),
$$
$$
\tilde \chi_1^{01}=-\left(h_{,1}^{01}- h^{11} \pi_{11}+ h^{00} \pi_{00}
\right),
$$
\begin{equation}
\label{26}\tilde \chi_1^{00}=-\left(h_{,1}^{00}-2 h^{01} \pi_{11}-
2 h^{00} \pi_{01}\right). 
\end{equation}

Conservation of the primary constraints in time leads to the secondary 
constraints
\begin{equation}
\label{27}\dot \Pi_1^{11}=\left\{ \Pi_1^{11},H\right\} = \tilde \chi_1^{11}, 
\dot \Pi_1^{01} =\left\{ \Pi_1^{01} ,H\right\} = \tilde \chi_1^{01}, 
\dot \Pi _1^{00} = \left\{ \Pi _1^{00},H\right\} = \tilde \chi _1^{00}. 
\end{equation}

All secondary constraints have zero PB with the primary constraints 
and among themselves have the following PB \footnote {Terms with derivatives 
must be carefully treated taking into account the distributional character of 
PB in the infinite dimensional case, that is, field theory 
\cite {Sund, KM2005}.} 
\begin{equation}
\label{28}\left\{ \tilde \chi_1^{01} ,\tilde \chi _1^{00}\right\} = \tilde 
\chi _1^{00},\left\{ \tilde \chi_1^{01} ,\tilde \chi_1^{11}\right\} =-\tilde
\chi_1^{11},\left\{ \tilde \chi_1^{11},\tilde \chi _1^{00}\right\} =2\tilde
\chi_1^{01}.
\end{equation}

The Hamiltonian (\ref {25}) is a linear combination of secondary constraints 
and, because of the PB of (\ref{28}), the Dirac canonical procedure is 
completed by the 
presence of six first class constraints for the six remaining canonical pairs 
(already three pairs $\xi_{\alpha \beta}^0$, $\Pi_{0}^{\alpha \beta}$ have 
been eliminated) resulting in there being zero degrees of freedom. The 
secondary constraints (\ref{26}) in Lagrangian language correspond to 
equations of motion obtained by varying such non-dynamical variables as those 
of (\ref {1.14}). These equations cannot be solved for the fields being varied 
and so they are not auxiliary fields. These equations correspond to first 
class constraints in Dirac language and are different in this respect from the 
algebraic constraints arising in Maxwell electrodynamics (\ref {1.17}). (The 
effect of 
performing a reduction by using the solution of first class constraints will 
be considered in the next section.)

In the Dirac procedure we have to first find all constraints, then eliminate 
the second class constraints. Only after these steps have been performed can we
discuss gauge fixing, etc. 

To find the full gauge invariance of the action using the Castellani procedure 
\cite {Castellani}, it is important to determine the complete set of first 
class constraints. If some of the first class constraints are solved, they 
will not be present in the gauge generator and the some of gauge symmetries 
cannot 
be restored. We can obtain, at most, partial gauge symmetries (e.g., only 
the spatial diffeomorphism) or even possibly the wrong gauge symmetries. 

The generator $G$ of gauge transformation, following Castellani 
\cite {Castellani}, is found by first setting 
$G_{\left( 1\right)}^a=C_P^a$ for the primary constraints 
($C_P^a=\left( \Pi_1^{11} ,\Pi_1^{01},\Pi_1^{00}\right) $) and then 
determining $G_{\left( 0\right) }^a\left( x\right)
=-\left\{ C_P^a,H_c\right\} \left( x\right) +\int dy\ \alpha _c^a\left(
x,y\right) C_P^c\left( y\right)$ where the functions $\alpha _c^a\left(
x,y\right) $ are found by requiring that $\left\{ G_{\left( 0\right)
}^a,H_c\right\} =0$. The full generator of gauge transformation is given by 
$$
G\left( \varepsilon ^a,\dot \varepsilon ^a\right) =\int
dx\left( \varepsilon
^a\left( x\right) G_{\left( 0\right) }^a\left( x\right) +\dot \varepsilon
^a\left( x\right) G_{\left( 1\right) }^a\left( x\right) \right). 
$$ 
In our case this leads to the following expression, using the three primary 
and three secondary first class constraints,
$$
G\left( \varepsilon \right) =\int dx\left[ \varepsilon \left( -\tilde 
\chi_1^{01} -\xi_{00}^1
\Pi _1^{00}+\xi_{11}^1\Pi_1^{11}\right) +\dot \varepsilon \Pi_1^{01} \right. 
$$
\begin{equation}
\label{29}\left. +\varepsilon _1\left( -\tilde \chi_1^{11}-2\xi_{01}^1
\Pi_1^{11} 
-2\xi_{00}^1 \Pi_1^{01}\right) +\dot \varepsilon _1\Pi_1^{11}+
\varepsilon ^1\left( -\tilde \chi _1^{00}+2\xi_1^{11}\Pi_1^{01} +2\xi_{01}^1 
\Pi _1^{00}\right) +\dot \varepsilon ^1\Pi _1^{00}\right]. 
\end{equation}

The PB of generators (\ref{29}) have a closed off-shell algebra similar to 
that of ordinary gauge theories: 
\begin{equation}
\label{30}\left\{ G\left( \varepsilon \right) ,G\left( \eta \right) \right\}
=G\left( \tau ^c=C^{cab}\varepsilon ^a\eta ^b)\right) 
\end{equation}
where $\varepsilon ^a=\left( \varepsilon ^1(\varepsilon ),\varepsilon
^2(\varepsilon _1),\varepsilon ^3(\varepsilon ^1)\right) $ and the only 
non-zero structure constants $C^{cab}$ are $C^{132}=2=-C^{123},$ 
$C^{212}=1=-C^{221},C^{331}=1=-C^{313}$. These reflect the structure of the 
algebra of the PB among the first class constraints. More explicitly, 
these relations are
$$
\tau=2\varepsilon^{1}\eta_{1}-2\varepsilon_{1}\eta^{1},
\tau_{1}=\varepsilon \eta_1-\varepsilon_1 \eta,
\tau^{1}=\varepsilon^{1}\eta-\varepsilon \eta^{1}.
$$

Now we can find the transformations for all fields appearing in the initial 
Lagrangian that follow from $\delta \left(field\right)=\left\{field,G\right\}$:
$$
\delta \xi_{11}^1=\dot \varepsilon_1-2 \varepsilon_1 \xi_{01}^1
+\varepsilon \xi_{11}^1,
$$
\begin{equation}
\label{31}\delta \xi_{01}^1=\frac 12 \dot \varepsilon- \varepsilon_1 
\xi_{00}^1+\varepsilon^1 \xi_{11}^1,
\end{equation}
$$
\delta \xi_{00}^1=\dot \varepsilon^1- \varepsilon \xi_{00}^1
+2 \varepsilon^1 \xi_{01}^1;
$$
$$
\delta h^{11}=-\varepsilon h^{11}-2 \varepsilon^1 h^{01},
$$
\begin{equation}
\label{32}\delta h^{01}=\varepsilon_1 h^{11}- \varepsilon^1 h^{00}, 
\end{equation}
$$
\delta h^{00}=\varepsilon h^{00}+2 \varepsilon_1 h^{01};
$$
$$
\delta \pi_{11}=\varepsilon_{1,1}-2 \varepsilon_1 \pi_{01}
+\varepsilon \pi_{11},
$$
\begin{equation}
\label{33}\delta \pi_{01}=\frac 12 \varepsilon_{,1}- \varepsilon_1 
\pi_{00}+\varepsilon^1 \pi_{11}, 
\end{equation}
$$
\delta \pi_{00}=\varepsilon_{,1}^1- \varepsilon \pi_{00}
+2\varepsilon^1 \pi_{01}.
$$

From (\ref {33}) and using the strong equalities (\ref {24}), we obtain
$$
\delta \xi_{11}^0= -\varepsilon_{1,1}-2 \varepsilon_1 \xi_{01}^0
+\varepsilon \xi_{11}^0,
$$
\begin{equation}
\label{34}\delta \xi_{01}^0=-\frac 12 \varepsilon_{,1}- \varepsilon_1 
\xi_{00}^0+\varepsilon^1 \xi_{11}^0,
\end{equation}
$$
\delta \xi_{00}^0=-\varepsilon_{,1}^1- \varepsilon \xi_{00}^0
+2\varepsilon^1 \xi_{01}^0.
$$

One can easily check the gauge invariance of the Lagrangian $\tilde L_{2}'$ of 
(\ref {19}) using the transformations of (\ref {31},\ref {32},\ref {34}). It is
\begin{equation}
\label{35}\delta \tilde L_2^{'}=\left(h^{11} \varepsilon_{1,1}+
h^{01} \varepsilon_{,1}+h^{00} \varepsilon_{,1}^1\right)_{,0}-
\left(h^{11} \dot \varepsilon_1+h^{01} \dot \varepsilon+
h^{00} \dot \varepsilon^1\right)_{,1} 
\end{equation}
and so $ \tilde L_2^{'}$ is invariant up to total derivatives. However, a 
variation of 
the total derivatives appearing in (\ref {18}) results in a contribution that 
{\it exactly} compensates (\ref {35}). Keeping the initial form $\tilde L_2$ 
of the Lagrangian, we have exact invariance under the transformations of 
(\ref {31}, \ref {32}, \ref {34}). This illustrates the importance of surface 
terms in retaining invariance of the Lagrangian and shows that the elimination 
of surface terms can affect its gauge invariance. (For a discussion of a 
similar occurrence in SUSY models, see \cite {KM2001}.) 

The transformations of (\ref {31}, \ref {32}, \ref {34}) can be written in a 
compact form which is similar to one appearing in \cite {Deser}\footnote {Due 
to the transfromations of \cite {Deser} we actually were able to recognize the 
possibility of recasting the change of variables (\ref {1.4}) into covariant 
form. These were found initially in component form in \cite {KKM1}. A similar 
change of variables is used for the antisymmetric part of the affine 
connection \cite {DDM1993} in generalized GR models formulated along the line 
of \cite {EIN1925}.} 
\begin{equation}
\label{36a}\delta h^{\alpha \beta}
=\left(\epsilon^{\alpha \lambda}h^{\sigma \beta}+
\epsilon^{\beta \lambda}h^{\sigma \alpha} \right)\zeta_{\lambda \sigma},
\end{equation}
\begin{equation}
\label{36b}\delta \xi_{\alpha \beta}^{\lambda}=-\epsilon^{\lambda \rho}
\zeta_{\alpha \beta , \rho}-\epsilon^{\rho \sigma}\left(\xi_{\alpha \rho}^
{\lambda}\zeta_{\beta \sigma}+\xi_{\beta \rho}^
{\lambda}\zeta_{\alpha \sigma} \right), 
\end{equation}
where $\epsilon^{\alpha \rho}$ is the antisymmetric tensor 
($\epsilon^{01}=1$) and
$\zeta_{\alpha \beta}$ is a symmetric tensor with components 
$\zeta_{00}=\varepsilon^{1}$, $\zeta_{11}=\varepsilon_{1}$,
$\zeta_{01}=\frac{1}{2}\varepsilon$. From (\ref {36b}) using (\ref {4},
\ref {1.4}) we obtain the transformation of $\Gamma_{\alpha \beta}^{\lambda}$ 
\begin{equation}
\label{36c}\delta \Gamma_{\alpha \beta}^{\lambda}=
-\frac {1}{2}\epsilon^{\lambda \rho}\zeta_{\alpha \beta , \rho}
+\delta_{\alpha}^{\lambda}\epsilon^{\nu \rho}\zeta_{\nu \beta, \rho}
\end{equation}
$$
-\epsilon^{\rho \sigma}\left[\Gamma_{\alpha \rho}^{\lambda}\zeta_{\beta \sigma}
-\frac {1}{2}\delta_{\rho}^{\lambda}\Gamma_{\alpha \nu}^{\nu} 
\zeta_{\beta \sigma}
-\delta_{\alpha}^{\lambda}\Gamma_{\beta \rho}^{\nu} 
\zeta_{\nu \sigma}
+\frac {1}{2}\delta_{\alpha}^{\lambda}\Gamma_{\rho \nu}^{\nu} 
\zeta_{\beta \sigma}
+\frac {1}{2}\delta_{\alpha}^{\lambda}\Gamma_{\beta \nu}^{\nu} 
\zeta_{\rho \sigma}
\right]
$$
$$
+\left(\alpha \leftrightarrow \beta \right).
$$

The Einstein form of the Lagrangian $L_{2}\left(h,\Gamma \right)$ is 
invariant under the transfromations of (\ref {36a}) and (\ref {36c}). 

All the usual canonical properties of non-Abelian gauge theories are present 
in the approach outlined here. We have local PB with field independent 
structure constants, a closed off-shell algebra of generators, and exact 
invariance of the Lagrangian under gauge transformations of the original 
fields. In \cite {KKM1}, only the surface term $-\left(h^{\alpha \beta} 
\xi _{\alpha \beta}^k\right)_{,k}$ was added to $\tilde L_2$ and so 
$\xi_{\alpha \beta}^{0}$ played role of a generalized coordinate \footnote 
{This, of 
course, cannot affect the result as the roles of coordinates 
and momenta are interchangable in the Hamiltonian formulation. The 
transformations in both cases are the same and only from a purely 
computational point of view might some preference exist.}.

\section {Lagrangian reduction based on solutions of first class 
constraints}

Again, starting with the same Lagrangian (\ref {19}), we find the equations of 
motion associated with the non-dynamical fields to obtain the Lagrangian 
constraints. These are identical to the PB of the primary constraints with the 
Hamiltonian (\ref {27}). We find that
\begin{equation}
\label{37}\frac {\delta \tilde L_2^{'}}{\delta \xi_{11}^1}
= h_{,1}^{11} - 2 h^{11}\xi_{01}^0 - 2 h^{01}\xi_{00}^0=0,  
\end{equation}
\begin{equation}
\label{38}\frac {\delta \tilde L_2^{'}}{\delta \xi_{01}^1}
= h_{,1}^{01} + h^{11} \xi_{11}^0 - h^{00}\xi_{00}^0=0,
\end{equation}
\begin{equation}
\label{39}\frac {\delta \tilde L_2^{'}}{\delta \xi_{00}^1}
= h_{,1}^{00} + 2 h^{01}\xi_{11}^0 + 2 h^{00}\xi_{01}^0=0. 
\end{equation}

These are the only three equations of motion out of nine in total that have no 
time derivatives. 

In the Dirac approach, we cannot solve them to eliminate any variables 
as they are all first class constraints. From the Lagrangian point of view, 
they are not equations for the auxiliary fields, since the fields being varied 
do not appear on right hand side of (\ref {37}-\ref {39}). If one were to use 
them anyway, as is done in ADM approach \cite {ADM2}, to solve for two  
variables algebraically, such as from (\ref {38}) and (\ref {39})
\begin{equation}
\label{40}\xi_{00}^0=\frac {1}{h^{00}}\left(  h_{,1}^{01} + h^{11} \xi_{11}^0 
\right),
\end{equation}
\begin{equation}
\label{41}\xi_{01}^0 =-\frac {1}{h^{00}}\left(\frac {1}{2}  h_{,1}^{00} 
+ h^{01}\xi_{11}^0\right), 
\end{equation}
and then substitute these equations  back into the original Lagrangian 
(\ref {19}) to eliminate $\xi_{00}^{0}$ and $\xi_{01}^{0}$, one finally 
obtains the reduced Lagrangian
\begin{equation}
\label{42}\tilde L_2^{'\left(1\right)}=-\dot h^{11} \xi_{11}^0 + 2 \dot h^{01} 
\frac {h^{01}}{h^{00}}\xi _{11}^0 - \dot h^{00}\frac {h^{11}}{h^{00}} 
\xi_{11}^0 -\xi_{11}^{1}\left( h_{,1}^{11}-2h_{,1}^{01}\frac {h^{01}}{h^{00}} 
+h_{,1}^{00}\frac {h^{11}}{h^{00}}\right)+S  
\end{equation}
where $S$ consists of two total derivatives   
\begin{equation}
\label{43}S=\left(\dot h^{01}\ln{h^{00}}\right)_{,1}
-\left(h_{,1}^{01}\ln{h^{00}}\right)_{,0}. 
\end{equation}
(This is similar to eq.(A.26) of \cite {Fad1982}.) Note that only two out of 
the three equations (\ref {37}-\ref {39}) can be solved because after 
substitution of (\ref {40},\ref {41}) into (\ref {37}) all the variables 
$\xi_{\alpha \beta}^{0}$ dissapear, converting equation (\ref {37}) into a 
differential constraint that is retained in the reduced Lagrangian 
(\ref {42}). After this reduction, we have five fields out of the original 
nine left in $\tilde L_2^{'(1)}$ in (\ref {42}) so that two equations of 
motion have served to eliminate four dynamical fields. This is a familiar 
feature of the Lagrangian reduction in $4D$ \cite {ADM2}; solving one first 
class constraint leads to the disappearance of two variables in the reduced 
Lagrangian. In \cite {ADM2}, after solving 30 equations, 34 variables 
disappear from the reduced Lagrangian (as four equations out of the 30 that 
have no time derivatives are first class constraints).

Moreover, if we perform variation of the reduced Lagrangian (\ref {42}) with 
respect to $\xi_{11}^{0}$, we obtain
\begin{equation}
\label{44}\frac {\delta \tilde L_2^{'(1)}}{\delta \xi_{11}^{0}}=
-\dot h^{11}  + 2 \dot h^{01} \frac {h^{01}}{h^{00}}- 
\dot h^{00}\frac {h^{11}}{h^{00}}.  
\end{equation}

The variation of the original Lagrangian (\ref {19}) with respect to the same 
variable gives
\begin{equation}
\label{45}\frac {\delta \tilde L_2^{'}}{\delta \xi_{11}^{0}}=
-\dot h^{11}  - 2 h^{11}\xi_{01}^{1}- h^{01}\xi_{00}^{1}.   
\end{equation}

The use of (\ref {40},\ref {41}) or any of the other equations of motion 
cannot account for the difference between (\ref {44}) and (\ref {45}). Thus, 
the reduced Lagrangian $\tilde L_2^{'(1)}$ is not equivalent to the original 
one $\tilde L_2^{'}$. If one, despite of this inconsistency, wants to find the 
Hamiltonian associated with this {\it new} Lagrangian (\ref {42}), the Dirac 
procedure must now be repeated. 

Introducing momenta conjugate to the five fields
\begin{equation}
\label{46}\Pi_{1}^{11}\left(\xi_{11}^{1}\right),
\Pi_{0}^{11}\left(\xi_{11}^{0}\right),
\pi_{\alpha \beta}\left(h^{\alpha \beta}\right)
\end{equation}
leads to five primary constraints
\begin{equation}
\label{47}\Pi_{1}^{11}\approx 0,\Pi_{0}^{11}\approx 0,\pi_{11}+\xi_{11}^{0}
\approx 0, 
\pi_{01}-\frac {h^{01}}{h^{00}}\xi_{11}^{0}\approx 0, 
\pi_{00}+\frac {h^{11}}{h^{00}}\xi_{11}^{0}\approx 0. 
\end{equation}

Among them we have a pair which are second class of a special form (the second 
and third constraints of (\ref {47})\footnote {We can equally take the second 
and fourth (or alternatively the second and fifth) constraints without 
affecting the final result.}) that allows us 
to set $\Pi_{0}^{11}=0$ and $\xi_{11}^{0}=-\pi_{11}$. Substitution of these  
equalities into the Hamiltonian and the remaining constraints gives the 
reduced total Hamiltonian
\begin{equation}
\label{48}H_{T}^{\left( 1 \right)}=H_{c}^{\left( 1 \right)}+
\lambda^{01}\left(\pi_{01}+\frac {h^{01}}{h^{00}}\pi_{11} \right)+
\lambda^{00}\left(\pi_{00}-\frac {h^{11}}{h^{00}}\pi_{11} \right)+
\Lambda_{1}^{11}\Pi_{11}^{1} 
\end{equation}
where
\begin{equation}
\label{49}H_{c}^{\left( 1 \right)}=-\xi_{11}^{1} \chi_{1}^{11}
\end{equation}
with
\begin{equation}
\label{49a}\chi_{1}^{11}=-h_{,1}^{11}+2h_{,1}^{01}\frac {h^{01}}{h^{00}} 
-h_{,1}^{00}\frac {h^{11}}{h^{00}}.
\end{equation}

Continuing with the Dirac procedure, we find the secondary constraints. The PB 
among all primary constraints are zero, the only non-obvious one being 
$\left\{\pi_{01}+\frac {h^{01}}{h^{00}}\pi_{11},
\pi_{00}-\frac {h^{11}}{h^{00}}\pi_{11}\right\}$.

The only secondary constraint that arises is  
\begin{equation}
\label{50}\dot \Pi_{1}^{11}=\left\{\Pi_{1}^{11},H_{c}^{\left( 1 \right)} 
\right\}=\chi_{1}^{11}, 
\end{equation}
showing that the Hamiltonian is a constraint. It is straightforward to show 
that the PB among all constraints, both primary and secondary, are zero so 
that 
we have four first class constraints for four pairs of canonical variables 
leaving us with no net degrees of freedom. At this point everything looks 
consistent as there are zero degrees of freedom, a local algebra of first 
class constraints and closure of the Dirac procedure. The Dirac constraint 
formalism applied to the reduced Lagrangian gives consistent results even 
though it is not equivalent to the original theory. As in the previous 
section, we can 
find a gauge transformation corresponding to this system of constraints. The 
resulting gauge generator is much simpler than that of (\ref {29}). It has 
though the same number of gauge parameters because the reduced total 
Hamiltonian (\ref {48}) also has three primary first class constraints. 
We find that
\begin{equation}
\label{51}G\left( \varepsilon \right) =\int dx\left[ 
\varepsilon^{1}\left( \pi_{01}+\frac {h^{01}}{h^{00}}\pi_{11} \right)
+\varepsilon\left(\pi_{00}-\frac {h^{11}}{h^{00}}\pi_{11} \right)
-\varepsilon_{1}\chi_{1}^{11}
+\dot \varepsilon_{1}\Pi_{1}^{11}\right].
\end{equation}

The algebra of this generator is closed even off-shell. This generator leads 
to the gauge transformations of the fields
\begin{equation}
\label{52}\delta \xi_{11}^{1}=\dot \varepsilon_{1}, \delta \Pi_{1}^{11}=0, 
\end{equation}
\begin{equation}
\label{53}\delta h^{11}=\varepsilon^{1}\frac {h^{01}}{h^{00}}-
\varepsilon \frac {h^{11}}{h^{00}}, \delta h^{01}=\frac {1}{2}\varepsilon^{1},
\delta h^{00}=\varepsilon, 
\end{equation}
\begin{equation}
\label{54}\delta \pi_{11}=\varepsilon \frac {1}{h^{00}}\pi_{11}+
\varepsilon_{1,1}-\varepsilon_{1}\frac{1}{h^{00}}h_{,1}^{00}.  
\end{equation}

To check the invariance of the reduced Lagrangian (\ref {42}), we also need 
the transformation of $\xi_{11}^{0}$ which can be easily restored by using the 
strong equality $\xi_{11}^{0}=-\pi_{11}$ so that
\begin{equation}
\label{55}\delta \xi_{11}^{0}=\varepsilon \frac {1}{h^{00}}\xi_{11}^{0}+
\varepsilon_{1,1}-\varepsilon_{1}\frac{1}{h^{00}}h_{,1}^{00}.  
\end{equation}

Using (\ref {52},\ref {53},\ref {55}), the variation of (\ref {42}) is
\begin{equation}
\label{56}\delta \tilde L_{2}^{'\left( 1 \right)}=
\left(\varepsilon \frac {1}{h^{00}}\xi_{11}^{1}-\varepsilon_{1}\frac 
{\dot h^{00}}{h^{00}}\right)
\frac {\delta \tilde L_{2}^{'\left( 1 \right)}}{\delta \xi_{11}^{1}}+S, 
\end{equation}
where $S$ is a term with total derivatives. Hence the Lagrangian is invariant 
only on-shell, which is a familiar feature of the ADM approach (the gauge 
generator in $4D$ has a closed algebra only on-shell \cite {Castellani}). 
In the previous section we were able to determine the transformations of all 
fields appearing in the original Lagrangian. It is not possible to do so now; 
by going back 
we can only restore the transformations of $\xi_{00}^{0}$ and $\xi_{01}^{0}$ 
using eqs. (\ref {40}) and (\ref {41}) but we cannot do this for  
$\xi_{01}^{1}$ and $\xi_{00}^{1}$ \footnote {Of course, after the Lagrangian 
reduction based on solution of first class constraints we cannot return to 
the variables $\Gamma$ because not all the transformations of the fields 
$\xi$ can be found. It is not difficult to repeat such a reduction directly in 
the Lagrangian when it is written in terms of $\Gamma$ and compare with the 
results of the previous section; they will be also different.}.  
Thus using a Lagrangian reduction which is 
based on employing solutions of the first class constraints leads to the 
gauge transformations for only some of the fields in the original 
Lagrangian and the Lagrangian is invariant only on-shell. This is to be 
compared with having exact invariance of the Lagrangian and gauge 
transformations for all the fields when one uses the Dirac approach to the 
Lagrangian before making a reduction based on all time independent equations 
of motion.

Moreover, if we consider the formulation of the $2D$ EH action using 
$g^{\alpha \beta}$ instead of $h^{\alpha \beta}$  as independent variables 
and perform the Lagrangian reduction using (\ref {38}) and (\ref {39}), 
the Lagrangian vanishes identically. To demonstrate this, we present the 
reduced Lagrangian (\ref {42}) in the following form
\begin{equation}
\label{57}\tilde L_{2}^{'}=-\left(h^{11}h^{00}-h^{01}h^{01}\right)_{,0}
\frac {1}{h^{00}}
\xi_{11}^{0}-\left(h^{11}h^{00}-h^{01}h^{01}\right)_{,1}\frac {1}{h^{00}}
\xi_{11}^{1}. 
\end{equation}

If we consider $g^{\alpha \beta}$ to be the independent variable, then
$h^{\alpha \beta}$ is just a short form for $\sqrt {-g}g^{\alpha \beta}$ and 
the particular combination that enters (\ref {57}) under derivatives is 
\begin{equation}
\label{58}h^{11}h^{00}-h^{01}h^{01}=-1, 
\end{equation}
and so, the Lagrangian (\ref {57}) vanishes identically.
However, Dirac analysis in this case leads to seven independent first class 
constraints, five of which are primary, and consequently there are five 
parameters characterizing the  group of gauge transformations \cite {KKM2}.

\section {Conclusion}

The second order form of the EH action (\ref {1}) in which the metric is the 
only independent dynamical field is invariant under a general coordinate 
transformations if all terms, including the terms with second order 
derivatives are present. It is possible to apply the standard Dirac canonical 
analysis and to keep simultaneously the effect of all terms when using an 
equivalent first order formulation. The oldest first order formulation which 
is the closest in form to the second order EH action is the affine-metric 
formulation of Einstein \cite {EIN1925}. This formulation when treated using 
the standard methods of quantum field theory should automatically retain the 
classical limit. Demonstration that such a limit exists in models based on new 
ideas constitutes a considerable problem in itself \cite {Nicolai}. 

The advantage of using this first order Einstein formulation of the action was 
recognized a long time ago by ADM  and was used by them as a starting point in 
their canonical analysis of GR \cite {ADM1,ADM2}. However, they did not 
apply the straightforward Dirac analysis and they performed a 
preliminary Lagrangian reduction using solutions of first class constraints 
\cite {ADM2} (see also \cite {Fad1982}). The reduced Lagrangian found in  
this way is not equivalent to the original Lagrangian and, to quote 
\cite {ADMlast}, ``does not represent the full statement of general 
relativity''. In the concluding remarks to the last paper of the ADM series 
\cite {ADMlast} (see also remarks in \cite {ADM3}), the authors suggested that 
in view of the many ambiguities that could arise in an attempt to quantize 
consistently at reduced level, it would seem more 
fruitful to return to the original Lagrangian (\ref {3}) and  
try to repeat reduction to the canonical form within the framework of quantum 
theory. To keep the gauge invariance of the original Lagrangian when  
quantizing, all first class constraints must be preserved and second class 
constraints can only be eliminated when they are of a special form, or used 
to modify the PB by passing to Dirac brackets.

The importance of preserving all first class constraints in the course of the 
Dirac quantization was analysed by Ashtekar and Horowitz \cite 
{AshtekarHorowitz1982}. They concluded that the Lagrangian reduced-space 
method is likely to yield an incomplete description of quantum gravity. (See 
also the subsequent discussions in \cite {DiracVsReduced}).

During the two decades between the last paper of the ADM series and the 
Ashtekar-Horowitz analysis, another result of \cite {ADMlast} 
got a lot of attention (e.g., see \cite {Kuchar}). This involves a 
geometrical interpretation of what remains after a Lagrangian reduction of 
field
variables. The Hamiltonian obtained from the {\it reduced} Lagrangian has 
been  emphasized stressing the geometrical significance of the 
reduced set of variables, leading also to a shift back from treating four-
dimensional Einstein spacetime to just treating space by itself and from 
constraint dynamics of the full 
Hamiltonian to geometrodynamics of a reduced Hamiltonian. 
The reduced Lagrangian is invariant only under {\it spatial} coordinate 
transformations \cite {DeWitt1967} and the disappearance of some symmetries is 
a strong indication of the inequivalence of the two approaches. The 
possibility of a $3D$ geometrical 
interpretation of the variables appearing in the reduced Lagrangian 
(corresponding to a particular slicing of spacetime) is a demonstration of the 
inconsistency of Lagrangian reduction because it contradicts the spirit of 
GR and furthermore introduces a restriction on the topology of spacetime 
\cite {Hawking}. This restriction originates in solving a part of the first 
class constraints. This `freezing' of symmetries is a sort of partial gauge 
fixing used in the ADM treatment \cite {ADM2} right from outset. However, the 
correct procedure is to fix the gauge only after the constraint analysis is 
performed \cite {Samuel2001,Nicolai}.  This approach also imposes the  
coordinate conditions that space-like surfaces remain space-like as is 
explicitly pointed out by Dirac \cite {Dirac1958}. This condition, which 
restricts the form of the general coordinate tranformations, obviously means 
abandoning four-dimensional spacetime symmetry as is 
clearly indicated in the conclusion of \cite {Dirac1958}.
 
Slicing and the imposition of coordinate conditions also contradict the 
canonical procedure, since for relativistic field theories a fundamental tenet 
of the canonical formulation is to not refer to the ambient spacetime \cite 
{IshamKuchar}. 
Any reference to a surface (or a particular subset of surfaces) already  
contradicts this, and it also implies the implicit introduction of 
extended objects into a local theory. It quite likely leads to 
non-locality, as models explicitly built using extended objects (such as 
string models) are essentially non-local by construction \cite {Woodard}. The 
common reference to similarities of {\it the hypersurface deformation 
algebra} (\ref {Dirac1}-\ref {Dirac3}) to string models as a sign of 
consistency of the constraint algebra of the reduced Hamiltonian is really a 
warning sign that GR, which is a local field theory, has somehow been 
converted into a non-local one. 

The Hamiltonian built from the reduced Lagrangian leads to the well-known 
non-local Dirac constraint algebra, which is difficult to quantize as it is 
not a true Lie algebra. There are also problems associated with defining time, 
finding physical observables, etc.; and as a result, numerous attempts to 
improve this approach by modifying the choice of variables 
\cite {newvariables}, reshuffling of the constraints 
\cite {Fotini1996, Thiemann2003}, etc. have been made.

However, once reduction has been performed any change of the {\it reduced} 
variables or reshuffling of the {\it reduced} constraints cannot cure these 
inconsistencies, as they are inherent to the framework of the reduced 
Lagrangian. The only possibility of resolving these problems is to not abandon 
the spirit of GR and the standard canonical procedure. 
The first order formulation of the EH action treated by the standard methods 
of constraint dynamics preserves all symmetries, as it does in ordinary gauge 
theories, and all results should be reconsidered prior to reduction. This was 
actually suggested by ADM (see \cite {ADMlast} Sec.7-8.1., ``Discussion of 
quantization'').

In \cite {KKM1,KKM2} an analysis of the $2D$ limit of the affine-metric first 
order formulation was performed without any {\it a priori} assumptions or 
restrictions such as those used in Lagrangian reduction, and the Dirac 
procedure 
was applied to see how some properties of GR that make it distinct from 
ordinary gauge theories might appear. However, it turned out that all 
properties of ordinary gauge theories remain manifestly intact if an 
alternative first order formulation is used based on a change of variables 
involving linear combinations of affine connections. This change of variables 
is easily generalized to any dimension by (\ref {1.4}) and has been employed 
in this article. The variables $\xi_{\alpha \beta}^{\lambda}$ of (\ref {1.4}) 
provide the alternative 
first order formulation of (\ref {1.5}) that considerably simplifies 
straightforward application of the Dirac procedure. (For details, see 
Appendix A.) The first steps of the Dirac algorithm (Sec.2) give results 
that are different from the ADM analysis which is based on the reduced 
Lagrangian. The origin of such difference lies in using solutions of equations 
of motion which correspond to first class constraints and the reduced 
Lagrangian obtained in this way is {\it not equivalent} to the original 
Lagrangian even at a classical level. (An example of where the Lagrangian 
reduction is equivalent to the Dirac approach is presented in Appendix C.)
The first class primary constraints obtained in the Dirac analysis 
(\ref {C20}) (Appendix C) cannot be eliminated since they constitute a 
first class subset 
with secondary constraints (\ref {1.16}) and any further constraint of higher 
order (e.g., tertiary constraints, either first or second class) 
cannot affect the first class character of this subset of constraints. 
Moreover, higher order constraints do not involve field variables conjugate to 
variables appearing in the primary constraints (\ref {C20}), i.e. the first 
class nature of primary constraints cannot be changed by occurence of higher 
order constraints. Consequently the Hamiltonian obtained by the 
Dirac approach cannot be reconciled with the Hamiltonian obtained from the 
reduced Lagrangian. The Dirac separation of constraints into first and second 
class is not merely a technical trick; these two classes of constraints are 
essentially different, as the first class constraints are an indication of the 
presence of gauge invariance. The knowledge of the gauge degrees of 
freedom is important when quantizing a model and must be kept in the 
formalism \cite {PlyushchayRazumov}. 

In Secs.3 and 4, we demonstrated by considering a simple example that Dirac 
reduction and Lagrangian reduction produce different results if Lagrangian 
constraints are solved without appropriate care. At first glance, the simple 
model which is treated in two different ways produces in both approaches the 
expected canonical results such as a local algebra of constraints with field 
independent structure constants, a closed off-shell algebra of gauge 
generators and the possibility of finding the gauge transformation of all 
fields. However, whereas the Dirac constraint analysis allows us to determine 
the gauge transformation of all fields from the original Lagrangian and to 
demonstrate exact gauge invariance of the Lagrangian, the Lagrangian reduction 
(based on eliminating variables by use of solutions of first class 
constraints) does not lead to well defined transformations of all the  
original fields and the reduced Lagrangian is invariant only on-shell. 

Moreover, if in more complicated cases the canonical analysis of the reduced 
Lagrangian leads to an algebra of constraints which is not a true Lie algebra, 
the problem of quantization arises but {\it this may be a problem of the 
reduced Lagrangian but it is not necessarily a problem of the original 
Lagrangian}.

In our simple $2D$ example we obtained quite different results using the two 
approaches showing that in the general case the canonical analysis of what 
is obtained after Lagrangian reduction can lead not only to gauge invariance 
on-shell but also, for example, to non-locality of PB, a closed algebra of 
generators only on-shell \cite{Castellani}, and consequently to a wrong or, 
at most, only partially right description of the initial Lagrangian.
We thus feel that for the EH action in higher dimensions it is natural to 
expect there to be an even more drastic deviations between the 
two approaches. What are they? We are not going to speculate here about all
possibilities, but we hope that we have been able to convince the reader that
the existing canonical formulation of the first order EH action has been 
obtained in 
a non-canonical way and its reduced, geometrodynamical, formulation is not 
equivalent to the original EH Lagrangian.
This gives rise to the very important question of what we are trying to 
quantize in canonical qravity. Is it the full Einstein GR theory or only the 
spatial, geometrodynamical, part of it?

The solution of the first class constraints in the first order formulation 
\cite {ADM2} is somehow related to a partial neglect of surface terms and 
imposing coordinate conditions in the second order analysis \cite {Dirac1958}.
It is natural to ask about the connection between solving first class 
constraints in first order formalism and breaking relativistic invariance 
by integrating out second order derivatives in the second order formulation. 
The full answer to this question can be given only if a generalization of the 
Dirac procedure is possible that allows us to deal directly with accelerations 
present in the second order 
EH action. Qualitatively, we expect that the term linear in acceleration 
corresponds to a primary constraint (see 
\cite {DuttDresden1986,GitmanTyutin1983,
high-derivatives}) which initiates a chain of higher order constraints. The 
elimination of such a term corresponds to cutting off the first term of a 
chain of constraints and has the same effect as solving a first class 
constraint in the first order formulation of the action. 

The numerous problems associated with canonical geometrodynamics are quite 
likely just problems of using the reduced Lagrangian (with its reduced 
symmetries), not an intrinsic characteristic of GR, and can actually be 
considered as an illustration of the fact that having only spatial symmetry is 
not enough for a consistent formulation of GR. The simplest and most natural 
possibility for resolving these problems has not been fully explored: instead 
of trying to improve the reduced formulation or attempting to find some new 
physics in the inconsistencies of canonical geometrodynamics, one should try 
to find a canonical formulation of GR by applying the Dirac procedure to  
its first order formulation without any {\it a priori} assumptions or 
restrictions. The use of an alternative first order formulation given in 
(\ref {1.5}) that is based on a generalization of the transformation found in 
the $2D$ limit of the action provides an example of how Dirac or 
Lagrangian reduction can be performed consistently. Possibly the use of these 
new variables is not sufficient to ensure a canonical form of GR that allows 
for quantization, and further modifications are needed in order to find a 
first 
order formulation that preserves all the properties of ordinary gauge theories 
in higher dimensions. In particular, we want to find a formulation that leads 
to a local algebra of constraints with field independent structure constants 
(as in \cite {KKM1,KKM2} and eqs.(\ref {1.16a}, \ref {28})).
The existence of such an algebra is needed to pass a crucial consistency test 
(for Hamiltonians) recently emphasized 
in \cite {Nicolai}. We believe that all possibilities have to be explored in 
this direction 
before any new physical hypothesis is introduced and before the question: 
``spacetime or space?'' (once answered by Einstein) can be posed again
\footnote {The answers to some more restricted questions such as: 
``Geometrodynamics: Spacetime or Space?'' is known \cite {Anderson}.}.

We would like to conclude our discussion by the epigraph to the second 
lecture, ``Geometrodynamics'', in the course on Canonical Quantization of 
Gravity at Banff Summer School \cite {Kuchar}:
 
{\it There is only the fight to recover
 
\hspace{3cm} what has been lost
 
And found and lost again and again

\hspace{3cm}T.S.Eliot: Four Quartets

\hspace{3cm}East Coker, 186-7.} 

\section{Acknowledgments}

The authors are greateful to D.G.C. McKeon for discussions during a few last 
months and for sharing his conversation with W. Kummer. 

We also would like to thank the organizers (especially, L. Smolin) and 
participants of Workshop ``Quantum Gravity in the Americas: Status and future 
directions'' (Perimeter Institute, October 2004) where many questions that we 
partially try to address in this article had arisen.

We thank E.V. Gorbar and S.R. Valluri for reading of the final version of 
manuscript, A. Buchel 
for pointing out Ref. \cite {Nicolai} and D. Matesich (SUNY) for making 
available to us preprint \cite {DuttDresden1986}.

\section{Appendix A}

Here the proof of the equivalence of the first order formulation 
$\tilde L_{d}\left(g,\xi \right)$ defined in (\ref {1.5}) with 
$L_{d}\left(g\right)$ of (\ref {1}) is presented for the case $d \neq 2$.

The variation of $\tilde L_{d}\left(g,\xi \right)$ with respect to 
$g^{\mu \nu}$ gives the standard result
\begin{equation}
\label{A1}\frac {\delta \tilde L_d}{\delta g^{\mu \nu}}=
\left(\sqrt {-g}\Delta_{\mu \nu}^{\alpha \beta}- 
\frac {1}{2}\sqrt {-g} g^{\alpha \beta}g_{\mu \nu}\right)R_{\alpha \beta}
\left(\xi \right) 
\end{equation}
where
\begin{equation}
\label{A2}R_{\alpha \beta}\left(\xi\right)= 
\xi _{\alpha \beta,\lambda }^\lambda -
\xi _{\alpha \sigma}^\lambda \xi_{\beta \lambda }^\sigma + 
\frac{1}{d-1} \xi _{\alpha \lambda }^\lambda \xi_{\beta \sigma }^\sigma.  
\end{equation}

From (\ref {A1}) it immediately follows that
\begin{equation}
\label{A3}R_{\mu \nu}-\frac {1}{2}g_{\mu \nu}R=0 
\end{equation}
or alternatively, by using the inverse of the expression multiplying  
$R_{\alpha \beta}$ in (\ref {A1})
\begin{equation}
\label{A4}\left(\sqrt {-g}\Delta_{\mu \nu}^{\gamma \sigma}- 
\frac {1}{2}\sqrt {-g} g^{\gamma \sigma}g_{\mu \nu}\right)^{-1}
=\frac {1}{\sqrt {-g}}\left(\Delta_{\gamma \sigma}^{\mu \nu}
-\frac {1}{d-2}g^{\mu \nu}g_{\gamma \sigma}\right)  
\end{equation}
(not defined in $2D$), we obtain
\begin{equation}
\label{A5}R_{\gamma \sigma}\left(\xi \right)=0. 
\end{equation}

$R_{\gamma \sigma}\left(\xi \right)$, as in the case when considering 
$L_{d}\left(g,\Gamma \right)$, now has to be expressed in terms of 
$g^{\alpha \beta}$. Varying $\tilde L_{d}$ with respect to 
$\xi_{\sigma \rho}^{\nu}$ we have
\begin{equation}
\label{A6}\frac {\delta \tilde L_d}{\delta \xi_{\sigma \rho}^{\nu}}
=-h_{,\nu}^{\sigma \rho} 
-h^{\mu \sigma}\xi_{\mu \nu}^{\rho}-h^{\mu \rho}\xi_{\mu \nu}^{\sigma}
-\frac {1}{d-1}
\left(h^{\mu \sigma}\xi_{\mu \lambda}^{\lambda}\delta_{\nu}^{\rho}+
h^{\mu \rho}\xi_{\mu \lambda}^{\lambda}\delta_{\nu}^{\sigma} \right). 
\end{equation}

This equation is easier to solve than the analogous equation for $\Gamma$. 
First, we can obtain the trace of $\xi_{\lambda \nu}^{\lambda}$. Multiplying 
(\ref {A6}) by $h_{\sigma \rho}$ (where $h_{\alpha \beta}h^{\beta \gamma}=
\delta_{\alpha}^{\gamma}$) we obtain
\begin{equation}
\label{A7}\xi_{\lambda \nu}^{\lambda}
=-\frac{1}{2}\frac{d-1}{d-2}h_{\gamma \tau}h_{,\nu}^{\gamma \tau}. 
\end{equation}

We see that here, as when using the variables $\Gamma$, $2D$ case is special.
Substitution of (\ref {A7}) into (\ref {A6}) gives
\begin{equation}
\label{A8} h^{\mu \sigma}\xi_{\mu \nu}^{\rho}
+h^{\mu \rho}\xi_{\mu \nu}^{\sigma}=D_{\nu}^{\sigma \rho}
\end{equation}
where
\begin{equation}
\label{A9}D_{\nu}^{\sigma \rho}=-h_{,\nu}^{\sigma \rho}
-\frac {1}{2\left(d-2\right)}\left(
h^{\mu \sigma}\delta_{\nu}^{\rho}+h^{\mu \rho}\delta_{\nu}^{\sigma}
\right)h_{\gamma \tau}h_{,\mu}^{\gamma \tau}.  
\end{equation}

Multiplying (\ref {A8}) by $h_{\alpha \rho}h_{\beta \sigma}$ and performing a 
permutation of the indices $\alpha, \beta ,\nu$ we obtain three equations; we 
add two (with the permutations $\left( \beta,\alpha,\nu\right)$ and 
$\left(\nu,\beta,\alpha \right)$) and subtract the third 
$\left(\alpha,\nu,\beta \right)$, so after multiplication by 
$\frac {1}{2}h^{\omega \beta}$ we obtain the solution for 
$\xi_{\nu \alpha}^{\omega}$ 
 \begin{equation}
\label{A10}\xi_{\nu \alpha}^{\omega}
=\frac {1}{2}\left[h_{\alpha \mu}D_{\nu}^{\mu \omega} 
+h_{\nu \mu}D_{\alpha}^{\mu \omega}
-h^{\omega \sigma}h_{\nu \mu}h_{\alpha \gamma}
D_{\sigma}^{\mu \gamma}\right]. 
\end{equation}

Substitution of (\ref {A10}) into (\ref {A3}) or (\ref {A5}) gives the 
Einstein equations for free space. (No reference to $\Gamma$ has been made.) 
Similarly, if we substitute this solution into the Lagrangian 
$\tilde L_{d}\left(g,\xi \right)$ of (\ref {1.5}), we obtain the reduced  
Lagrangian which is equivalent to the second order form of EH action  
$L_{d}\left(g\right)$ (including terms with second order derivatives).

The appearance of explicit dimensional dependence in (\ref {A7},\ref {A9}) 
seems to be inconsistent with using the Christoffel symbol\footnote {The 
expression for it has the coefficient $\frac {1}{2}$ in any dimension.}. To 
resolve this, let us consider the trace of $\Gamma$ expressed in terms of 
$\xi$. Using (\ref {1.4}) we find
\begin{equation}
\label{A11}\Gamma_{\nu \lambda}^{\lambda}
=-\frac{2}{d-1}\xi_{\nu \lambda}^{\lambda}, 
\end{equation}
so that upon substitution of (\ref {A7}) into (\ref {A11}) and remembering 
that $h^{\alpha \beta}$ is only short for $\sqrt {-g}g^{\alpha \beta}$, we 
obtain
\begin{equation}
\label{A12}\Gamma_{\nu \lambda}^{\lambda}
=\frac{1}{d-2}h_{\alpha \beta}h_{,\nu}^{\alpha \beta}=\frac{1}{d-2}\frac {1}
{\sqrt {-g}}g_{\alpha \beta}\left( \sqrt{-g}g^{\alpha \beta}\right)_{,\nu}=
-\frac{1}{2}g_{\alpha \beta}g_{,\nu}^{\alpha \beta} 
\end{equation}
which is a well-known expression. Similarly, the general case for arbitrary 
$\Gamma_{\alpha \beta}^{\gamma}$ can be demonstrated using (\ref {1.4}) and 
(\ref {A10}).

\section{Appendix B}

As an illustration of Dirac reduction (used in Sections 2-4) that employs  
elimination of only second class constraints that are of a special form, we 
prove the equivalence of the first and second order formulation of Maxwell 
electrodynamics at the level of the Hamiltonian. The first order form of the 
Maxwell Lagrangian is
\begin{equation}
\label{B1}L_{M}=-\frac 12\left( \partial _\mu A_\nu -\partial _\nu A_\mu 
\right) F^{\mu \nu }+\frac 14F_{\mu \nu }F^{\mu \nu }, 
\end{equation}
where $A_\mu $ and $F_{\mu \nu }=-F_{\nu \mu }$ are treated as independent 
fields. This formulation is equivalent to the standard second order form. This 
is 
obvious at the Lagrangian level, as the auxiliary field $F_{\mu \nu }$ can be 
easily eliminated. It is also not difficult to prove this at the Hamiltonian 
level by reducing the Hamiltonian that corresponds to (\ref {B1}) to the 
standard one by using the Dirac procedure.

Introducing momenta conjugate to all fields $\pi^{\mu} \left(A_{\mu}\right)$, 
$\Pi_{\mu \nu} \left(F^{\mu \nu} \right)$ we obtain
\begin{equation}
\label{B2}\pi ^\mu =\frac{\delta L}{\delta \left( 
\partial _0 A_{\mu }\right) }=F^{\mu 0} 
\end{equation}
and
\begin{equation}
\label{B3}\Pi _{\mu \nu }=\frac{\delta L}{\delta \left( 
\partial _0F^{\mu \nu }\right) }=0. 
\end{equation}

In $4D$, equations (\ref {B2}, \ref {B3}) give ten primary constraints
\begin{equation}
\label{B4}\phi ^\mu =\pi ^\mu -F^{\mu 0}\approx 0,\Phi _{\mu \nu }
=\Pi _{\mu \nu }\approx 0 
\end{equation}
and the total Hamiltonian is
\begin{equation}
\label{B5}H_p=H_c+\lambda _\mu \phi ^\mu +\Lambda ^{\mu \nu }\Phi _{\mu \nu } 
\end{equation}
where $\lambda _\mu ,\Lambda ^{\mu \nu }$ are Lagrange multipliers and
\begin{equation}
\label{B6}H_{c}=\partial _kA_0 F^{k0}+\partial _kA_mF^{km}
-\frac 12 F_{k0}F^{k0}-\frac 14F_{km}F^{km}. 
\end{equation}

The proof that the Dirac procedure closes can be found in \cite {Sund}. The 
first order formulation produces 14 constraints (2 first class and 12 
second class). However, we will proceed differently by eliminating step by 
step the second class constraints that are of a special form.

The non-zero fundamental PB are
\begin{equation}
\label{B7}\left\{ A_\mu ,\pi ^\nu \right\} =\delta _\mu ^\nu ,
\left\{ F^{\rho \sigma },\Pi _{\mu \nu }\right\} =\frac 12\left( 
\delta _\mu ^\rho \delta_\nu ^\sigma -\delta _\mu ^\sigma \delta _\nu ^\rho 
\right).  
\end{equation}

It is not difficult to calculate the PB among primary constraints. The only 
non-zero brackets are
\begin{equation}
\label{B8}\left\{ \phi ^\mu ,\Phi _{\rho \sigma }\right\} =-\frac 12\left( 
\delta _\rho ^\mu \delta _\sigma^0-\delta _\rho^0\delta _\sigma ^\mu \right),  
\end{equation}
so that there is a second class subset of primary constraints because
\begin{equation}
\label{B9}\left\{ \pi ^k-F^{k0},\Pi _{0m}\right\} =\frac 12\delta_{m}^{k}. 
\end{equation}

This subset is of a special form and allows one to solve these constraints 
leading to the reduced Hamiltonian (with a reduced number of variables) after 
substitution of
\begin{equation}
\label{B10}\Pi _{0k}=0, F^{k0}=\pi ^k 
\end{equation}
into the original Hamiltonian as well as into the remaining constraints. After 
the first stage of reduction, we have
\begin{equation}
\label{B11}H_p^{\left( 1\right) }=H_c^{\left( 1\right) }+\lambda _0\phi ^0
+\Lambda^{km}\Phi _{km} 
\end{equation}
where
\begin{equation}
\label{B12}H_c^{\left( 1\right) }=\partial _kA_0\pi ^k+\partial _kA_mF^{km}
-\frac 12\pi _k\pi^k-\frac 14F_{km}F^{km}.  
\end{equation}

We have now only seven independent fields and four primary constraints
\begin{equation}
\label{B13}\phi ^0=\pi ^0\approx 0,\Phi _{km}=\Pi _{km}\approx 0. 
\end{equation}

They have zero PB among themselves and conservation of these constraints now 
has to be considered. The conservation of the primary constraints gives rise 
to the secondary constraints $\chi ^0$ and $\chi_{km}$
\begin{equation}
\label{B14}\dot \pi ^0=\left\{ \pi ^0,H_c^{\left( 1\right) }\right\} 
=\left\{ \pi^0,-A_0\partial _k\pi ^k\right\} =\partial _k\pi ^k=\chi ^0 
\end{equation}
and
\begin{equation}
\label{B15}\dot \Phi _{km}=\left\{ \Pi _{km},H_c^{\left( 1\right) }\right\} 
=-\frac 12\left( \partial _kA_m-\partial _mA_k\right) +\frac 12F_{km}
=\chi_{km}. 
\end{equation}

It is obvious that the constraints $\chi_{km}$ constitute a second class 
subset of a special form with $\Phi _{km}$ and, as was done at the previous 
stage, can be eliminated by solving them for $\Pi_{km}$ and $F^{km}$
\begin{equation}
\label{B16}\Pi _{km}=0,F_{km}
=\left( \partial _kA_m-\partial _mA_k\right).  
\end{equation}

Upon substitution of (\ref {B16}) into $H_p^{\left( 1\right) }$, we obtain
\begin{equation}
\label{B17}H_p^{\left( 2\right) }=H_c^{\left( 2\right) }+\lambda _0\phi ^0 
\end{equation}
with
\begin{equation}
\label{B18}H_c^{\left( 2\right) }=\partial _kA_0\pi ^k-\frac 12\pi _k\pi ^k
+\frac 14\left(\partial _kA_m-\partial _mA_k\right) \left( \partial ^kA^m
-\partial^mA^k\right)  
\end{equation}
which is exactly the standard Hamiltonian in the second order formulation with 
only four fields and two first class constraints. This completes the proof of 
the equivalence between the two types of reduction at the pure Hamiltonian 
level. It is important to note that here we never `solve' first 
class constraints, unlike what occurs in the EH action when treated using the 
ADM formalism and so the problem associated with ADM reduction do not arise.

\section{Appendix C}

We perform Lagrangian reduction of the action of (\ref {1.5}) in a way that 
is consistent with the Dirac procedure by eliminating only non-dynamical 
fields by solving equations of motion with respect to fields used in the 
variation that leads to these equations. The Lagrangian (\ref {1.5}), after a 
complete separation of spatial and temporal components, can be presented as a 
sum of terms
\begin{equation}
\label{C1}L_{d}=h^{\alpha \beta}\dot \xi_{\alpha \beta}^{0}
+L_{1}\left(\xi_{\alpha \beta}^{0} \right)
+L_{2}\left(\xi_{00}^{k};\xi_{\alpha \beta}^{0}\right)
+L_{3}\left(\xi_{mn}^{k};\xi_{\alpha \beta}^{0} \right)
+L_{4}\left(\xi_{0m}^{k};\xi_{\alpha \beta}^{0},\xi_{mn}^{k}\right)
\end{equation}
where the purely dynamical part is
\begin{equation}
\label{C2}L_{1}=-\frac {d-2}{d-1}\left(h^{km}\xi_{0k}^{0}\xi_{0m}^{0}
+2h^{0k}\xi_{0k}^{0}\xi_{00}^{0}+h^{00}\xi_{00}^{0}\xi_{00}^{0}\right).
\end{equation}

The term with the non-dynamical field $\xi_{00}^{k}$ is
\begin{equation}
\label{C3}L_{2}=h^{00}\xi_{00,k}^{k}-2\xi_{00}^{k}\left(h^{00}\xi_{0k}^{0}
+h^{0p}\xi_{pk}^{0} \right),
\end{equation}
while the term with the non-dynamical field $\xi_{mn}^{k}$ is
\begin{equation}
\label{C4}L_3=h^{mn}\xi_{mn,k}^{k}-h^{km}\xi_{kq}^{p}\xi_{mp}^{q}
+\frac {1}{d-1}h^{km}\xi_{kp}^{p}\xi_{mq}^{q}
+\frac {2}{d-1}\left(h^{mk}\xi_{0k}^{0}+h^{0m}\xi_{00}^{0}\right)\xi_{mq}^{q},
\end{equation}
and finally the term that is at least linear in non-dynamical field 
$\xi_{0m}^{k}$ is
\begin{equation}
\label{C5}L_{4}=2h^{0m}\xi_{0m,k}^{k}-h^{00}\xi_{0m}^{k}\xi_{0k}^{m}
+\frac {1}{d-1}h^{00}\xi_{0p}^{p}\xi_{0q}^{q}
-2h^{0k}\xi_{0q}^{p}\xi_{pk}^{q}+\frac {2}{d-1}h^{0k}\xi_{0p}^{p}\xi_{kq}^{q}
\end{equation}
$$
-2h^{0k}\xi_{0k}^{p}\xi_{0p}^{0}+\frac {2}{d-1}h^{0k}\xi_{0p}^{p}\xi_{0k}^{0}
-2h^{km}\xi_{0m}^{p}\xi_{pk}^{0}+\frac {2}{d-1}h^{00}\xi_{0p}^{p}\xi_{00}^{0}.
$$

The $2D$ limit of (\ref {C1}) is obtained by setting $d=2$ and putting all 
spatial indices equal to one, giving (\ref {19}). 

We have three ``non-dynamical'' fields among the fields $\xi$ (i.e., fields 
that enter 
the Lagrangian without any time derivatives): $\xi_{00}^{k}$, $\xi_{0m}^{k}$ 
and $\xi_{mn}^{k}$. The first field enters only linearly and cannot be 
eliminated. (This term (\ref {C3}) corresponds to a first class constraint in 
the Dirac approach.) 
Variation of (\ref {C5}) with respect to $\xi_{0b}^{a}$ gives
\begin{equation}
\label{C6}-2h^{00}\xi_{0a}^{b}+\frac {2}{d-1}h^{00}\xi_{0c}^{c}\delta_{a}^{b}=
D_{a}^{0b}\left(h^{\alpha \beta},\xi_{\alpha \beta}^{0},\xi_{mn}^{k} 
\right). 
\end{equation}

The left side of this equation is not invertible and not all components can be 
eliminated because the trace of the left side is zero. To preserve the 
tensorial character of variables in the reduced Lagrangian it is better to 
introduce an extra (pure auxiliary) field $\theta$ by performing a change of 
variables in the following term
\begin{equation}
\label{C7}\frac {1}{d-1} h^{00}\xi_{0p}^{p}\xi_{0q}^{q}=
\frac {1}{d-1} h^{00}\xi_{0p}^{p}\theta-\frac {1}{4}\frac {1}{d-1} h^{00}
\theta\theta.
\end{equation}

Introducing of this extra field $\theta$ 
allows us to solve (\ref {C6}) for all components of $\xi_{0m}^{k}$. Variation 
of (\ref {C5}), taking into account (\ref {C7}), gives
\begin{equation}
\label{C8}\xi_{0a}^{b}=\frac {1}{h^{00}}\left[-h_{,a}^{0b}-h^{0k}\xi_{ka}^{b}
-h^{0b}\xi_{0a}^{0}-h^{kb}\xi_{ka}^{0}+\frac {1}{d-1}\delta_{a}^{b}
\left(\frac {1}{2}h^{00}\theta
+h^{0k}\xi_{kq}^{q}+h^{0k}\xi_{0k}^{0}+h^{00}\xi_{00}^{0}  \right) \right].
\end{equation}

Substitution of this leads to the reduced Lagrangian (with the 
$\left(d-1\right)^2$ components of $\xi_{0m}^{k}$ completely eliminated)
\begin{equation}
\label{C9}L_{d}^{\left(1\right)}=h^{\alpha \beta}\dot \xi_{\alpha \beta}^{0}
+L_{1}^{\left(1\right)}\left(\xi_{\alpha \beta}^{0} \right)
+L_{2}^{\left(1\right)}\left(\xi_{00}^{k};\xi_{\alpha \beta}^{0}\right)
+L_{3}^{\left(1\right)}\left(\theta;\xi_{\alpha \beta}^{0} \right)
+L_{4}^{\left(1\right)}\left(\xi_{nm}^{k};\xi_{\alpha \beta}^{0}\right)
\end{equation}
where now
$$
L_{1}^{\left(1\right)}=L_{1}-2h^{0m}\left[\frac {1}{h^{00}}
\left(h_{,m}^{0k}+h^{0k}\xi_{0m}^{0}+h^{nk}\xi_{nm}^{0}
\right)\right]_{,k}
+\frac {2}{d-1}h^{0m}\left[\frac {1}{h^{00}}
\left(h^{0k}\xi_{0k}^{0}+h^{00}\xi_{00}^{0}\right)\right]_{,m}
$$
\begin{equation}
\label{C10}-\frac {1}{h^{00}}
\left(h_{,m}^{0k}+h^{0k}\xi_{0m}^{0}+h^{nk}\xi_{nm}^{0}\right)
\left(h_{,k}^{0m}-h^{0m}\xi_{0k}^{0}-h^{pm}\xi_{pk}^{0}\right)
\end{equation}
$$
+\frac {1}{d-1}\frac {1}{h^{00}}\left(h^{0m}\xi_{0m}^{0}
+h^{00}\xi_{00}^{0}\right)\left(h^{00}\xi_{00}^{0}-h^{0n}\xi_{0n}^{0}
-2h^{kn}\xi_{kn}^{0}\right);
$$
\begin{equation}
\label{C11}L_{2}^{\left(1\right)}=L_{2};
\end{equation}
\begin{equation}
\label{C12}L_{3}^{\left(1\right)}=\frac {1}{d-1}\left(h^{0m}\theta_{,m}
-\theta h^{km}\xi_{km}^{0}+\theta h^{00}\xi_{00}^{0} \right);
\end{equation}
$$
L_{4}^{\left(1\right)}=h^{mn}\xi_{mn,k}^{k}
-e^{km}\xi_{kq}^{p}\xi_{mp}^{q}
+\frac {1}{d-1}e^{km}\xi_{kp}^{p}\xi_{mq}^{q}
-2h^{0m}\left(\frac {h^{0p}}{h^{00}}\xi_{pm}^{k}\right)_{,k}
-\frac {2}{d-1}h^{0m}\left(\frac {h^{0p}}{h^{00}}\xi_{pq}^{q}\right)_{,m}
$$
\begin{equation}
\label{C13}+2\frac {h^{0k}h^{0m}}{h^{00}}\xi_{0p}^{0}\xi_{mk}^{p}
-\frac {2}{d-1}\frac {h^{0k}h^{0m}}{h^{00}}\xi_{0k}^{0}\xi_{mq}^{q}
+2\frac {h^{km}h^{0n}}{h^{00}}\xi_{kp}^{0}\xi_{nm}^{p}
\end{equation}
$$
-\frac {2}{d-1}\frac {h^{km}h^{0n}}{h^{00}}\xi_{km}^{0}\xi_{nq}^{q}
+\frac {2}{d-1}\left(h^{mk}\xi_{0k}^{0}+h^{0m}\xi_{00}^{0}\right)\xi_{mq}^{q}.
$$

Note, that the terms quadratic in $\theta$ have cancelled out and that the 
terms linear in $\theta$ lead to the additional first class constraint 
(\ref {C12}). The terms quadratic in $\xi_{mn}^{k}$ are
multiplied by 
\begin{equation}
\label{C14}e^{km}=h^{km}-\frac {h^{0k}h^{0m}}{h^{00}}
\end{equation}
where $e^{km}$ has the property: $e^{km}h_{mn}=\delta_{n}^{k}$. We can now 
reduce the 
Lagrangian  further by eliminating $\xi_{km}^{n}$. The variation of 
(\ref {C13}) with respect to $\xi_{bc}^{a}$ gives
\begin{equation}
\label{C15}e^{kb}\xi_{ka}^{c}+e^{kc}\xi_{ka}^{b}-\frac {1}{d-1}\xi_{kp}^{p}
\left(e^{kb}\delta_{a}^{c}+e^{kc}\delta_{a}^{b} \right)=D_{a}^{bc}
\end{equation}
where
\begin{equation}
\label{C16}D_{a}^{bc}\equiv -\frac {1}{2}h_{,a}^{bc}
+\frac {h^{0b}}{h^{00}}h_{,a}^{0c}
+\frac {h^{0b}h^{0c}}{h^{00}}\xi_{0a}^{0}
+\frac {h^{kb}h^{0c}}{h^{00}}\xi_{ka}^{0}
\end{equation}
$$
+\frac {1}{d-1}e^{kb}\xi_{0k}^{0}\delta_{a}^{c}-\frac {1}{d-1}\frac {h^{0b}}
{h^{00}}\left(h_{,m}^{0m}+h^{km}\xi_{km}^{0}-h^{00}\xi_{00}
^{0}\right)\delta_{a}^{c}+\left(b \leftrightarrow c\right).
$$

The solution of (\ref {C15}) is similar to solution of (\ref {A6}) appearing 
in Appendix A. Multiplying (\ref {C15}) by $h_{bc}$, we obtain the trace
\begin{equation}
\label{C17}\xi_{ka}^{k}=\frac 12 \frac {d-1}{d-2}h_{bc}D_{a}^{bc}
\end{equation}
and substituting of (\ref {C17}) into (\ref {C15}) leads to
\begin{equation}
\label{C18}e^{kb}\xi_{ka}^{c}+e^{kc}\xi_{ka}^{b}=D_{a}^{bc}+\frac 12 
\frac {1}{d-2}h_{pq}D_{k}^{pq}\left(e^{kb}\delta_{a}^{c}
+e^{kc}\delta_{a}^{b}\right)\equiv \tilde D_{a}^{bc}. 
\end{equation}

Multiplying (\ref {C18}) by $h_{rb}h_{sc}$ and performing a permutation of 
the indices $r, s ,a$ (as in (\ref {A8})) and then multiplication 
by $\frac {1}{2}e^{ns}$, we obtain the solution for $\xi_{ar}^{n}$ 
\begin{equation}
\label{C19}\xi_{ar}^{n}=\frac {1}{2}\left[h_{rb}\tilde D_{a}^{bn} 
+h_{ab}\tilde D_{r}^{bn}-e^{ns}h_{ab}h_{rc}
\tilde D_{s}^{bc}\right] 
\end{equation}
or in terms of (\ref {C16})
\begin{equation}
\label{C19a}\xi_{ar}^{n}=\frac {1}{2}\left[h_{rb}D_{a}^{bn} 
+h_{ab}D_{r}^{bn}-e^{ns}h_{ab}h_{rc}D_{s}^{bc}+\frac {1}{d-2}h_{pq}
D_{k}^{pq}h_{ar}e^{kn}\right]. 
\end{equation}

Substitution of (\ref {C19a}) back into the Lagrangian 
$L_{d}^{\left(1\right)}$ 
produces the reduced 
Lagrangian with the non-dynamical fields $\xi_{mn}^{k}$ all absent. We can use 
this reduced Lagrangian to pass to a Hamiltonian formulation which is 
different from the ADM-reduced formulation, as all components of 
$\xi_{\alpha \beta}^{0}$ 
are still present (and not only its spatial components). This should lead, in 
principle, to a restoration of full gauge invariance that will involve all 
components of $g^{\alpha \beta}$. The solution of the equations of motion 
corresponding to the first class constraints in the Dirac approach leads  to a 
non-equivalent reduced Lagrangian with a loss of the possibility of restoring 
full gauge invariance of the initial Lagrangian. This is illustrated for $d=2$ 
in Sec.4. We note that our elimination of the non-dynamical variables 
produces an alternative formulation of the Einstein-Hilbert action that is 
linear in time derivatives of the dynamical fields and well suited for 
application of the standard Dirac procedure.

The Dirac analysis applied to the reduced Lagrangian $L_{d}^{\left(1\right)}$ 
gives (as in Sec.2) the second class primary constraints
$$
\pi _{\alpha \beta} \approx 0,
\Pi_0^{\alpha \beta} - \sqrt{-g} g^{\alpha \beta} \approx 0
$$
as well as two first class primary constraints
\begin{equation}
\label{C20}\Pi_{k}^{00} \approx 0, \pi \approx 0,
\end{equation}
where $\Pi_{k}^{00},\pi$ are momenta conjugate to fields $\xi_{00}^{k},
\theta$. Conservation of the constraints of (\ref {C20}) in time 
(using (\ref {C3}, \ref{C12})) leads to secondary constraints which are 
equivalent to (\ref {1.16}).


\begin{thebibliography}{99}

\bibitem{Nicolai}H. Nicolai, K. Peeters and M. Zamaklar, hep-th/0501114.

\bibitem{Bergmann1951}J.L. Anderson and P.G. Bergmann, Phys. Rev. {\bf 83} 
(1951) 1018.

\bibitem{Dirac1950}P.A.M. Dirac, Can. J. Math. {\bf 2} (1950) 129.

\bibitem{Dirac1951}P.A.M. Dirac, Can. J. Math. {\bf 3} (1951) 1.

\bibitem{PiraniSchild1950}F.A.E. Pirani and A. Schild, Phys. Rev. {\bf 79} 
(1950) 986.

\bibitem{gamma-gamma}F.A.E. Pirani, A. Schild and S. Skinner, Phys. Rev. 
{\bf 87} (1952) 452.

\bibitem{Dirac1958}P.A.M. Dirac, Proc. Roy. Soc. {\bf A246} (1958) 333.

\bibitem{Dirac1959}P.A.M. Dirac, Phys. Rev. {\bf 114} (1959) 924.
 
\bibitem{Dirac1964}P.A.M. Dirac, {\it Lectures on Quantum Mechanics} (Belfer 
Graduate School of Sciences, Yeshiva University, New York, 1964).

\bibitem{LL}L.D. Landau and E.M. Lifshits, {\it The Classical Theory of 
Fields}, 4th edition (Pergamon Press, Oxford, 1975).

\bibitem{RT1974}T. Regge and C. Teitelboim, Ann. Phys. {\bf 88} (1974) 286; 
V.O. Soloviev, Phys. Lett. {\bf B292} (1992) 30.

\bibitem{Padmanabhan2004}T. Padmanabhan, gr-qc/0409089.

\bibitem{ADM4}R. Arnowitt, S. Deser and C.W. Misner, J. Math. Phys. {\bf 1} 
(1960) 434.

\bibitem{ABFKM2005}E. Anderson, J. Barbour, B.Z. Foster, B. Kelleher and 
N.O. Murchadha, Class. Quant. Grav. {\bf 22} (2005) 1795.

\bibitem{Thiemann2003LNP}T. Thiemann, Lectures on loop quantum gravity, in 
{\it Quantum Gravity}, edited by D. Guilini, C. Kiefer and C. L\"ammerzahl, 
Lecture Notes in Physics (Springer, Berlin, 2003) 39.

\bibitem{Ostrogradskii}M. Ostrogradsky, Memoires de l'Academie Imperiale des 
Science de Saint-Petersbourg, {\bf IV} (1850) 385.

\bibitem{GitmanTyutin1983}D.M. Gitman and I.V. Tyutin, Izvestiya Vuz. Fizika, 
{\bf 26} (1983) 61; Sov. Phys. J. (1984) 730

\bibitem{high-derivatives}I.L. Buchbinder and S.L. Lyakhovich, Izvestiya Vuz.
Fizika, {\bf 28} (1985); Sov. Phys. J. (1986) 746; C. Batlle, J. Gomis, J.M. 
Pons and N. Roman-Roy, J. Phys. A: Math. Gen. {\bf 21} (1988) 2693; 
V.V. Nesterenko, J. Phys. A: Math. Gen. {\bf 22} (1989) 1673.  

\bibitem{DuttDresden1986}S.K. Dutt and M. Dresden, Pure gravity as a 
constrainted second-order system, Preprint ITP-SB-86-32. 

\bibitem{EIN1925}A. Einstein, Sitzungsber.preuss. Akad. Wiss., phys.-math. 
{\bf K1}, (1925) 414 and {\it The complete collection of scientific papers}
(Nauka, Moskow, 1966), v.2, p.171.

\bibitem{EINEdd}A. Einstein, Sitzungsber.preuss. Akad. Wiss., phys.-math. 
{\bf K1}, (1923) pp. 32, 76, 137 and {\it The complete collection of 
scientific papers} (Nauka, Moskow, 1966), v.2, pp. 134, 142, 145.

\bibitem{DDM1993}T. Damour, S. Deser and J. McCarthy, Phys. Rev. {\bf D47} 
(1993) 1541.

\bibitem{Wald}R.M. Wald, {\it General Relativity} (The University of Chicago 
Press, Chicago, 1989), Appendix E.

\bibitem{ADM1}R. Arnowitt and S. Deser, Phys. Rev. {\bf 113} (1959) 745.

\bibitem{ADM2}R. Arnowitt, S. Deser and C.W. Misner, Phys. Rev. {\bf 116} 
(1959) 1322.

\bibitem{Sund}K. Sundermayer, {\it Constrained Dynamics}, 
Lecture Notes in Physics, {\bf 169} (Springer-Verlag, Berlin, 1982).

\bibitem{Mann}J. Gegenberg, P.F. Kelly, R.B. Mann and D. Vincent, Phys. Rev. 
{\bf D37} (1988) 3463; U. Lindstr\"om and M. Ro\v{c}ek, Class. Quant. Grav. 
{\bf 4} (1987) L79; S. Deser, gr-qc/9512022. 

\bibitem{2Dmodels}D. Grumiller, W. Kummer and D.V. Vassilevich, Phys. Rep.
{\bf 369} (2002) 327.

\bibitem{Thiemann2001}T. Thiemann, gr-qc/0110034.

\bibitem{KKM1}N. Kiriushcheva, S.V. Kuzmin and D.G.C. McKeon, Mod. Phys. Lett. 
{\bf A20} (2005) 1895: hep-th/0501204.

\bibitem{Hawking}S.W. Hawking, The path-integral approach to quantum gravity, 
in: {\it General Relativity. An Einstein centenary survey}. Eds. S.W. 
Hawking and W. Israel (Cambridge University Press, 1979), 746.

\bibitem{Kuchar}K. Kucha\v{r}, Canonical Quantization of Gravity, in 
{\it Relativity, Astrophysics and Cosmology}, edited W. Israel (D. Reidel Publ.
Comp, Dordrecht, 1972) 237.

\bibitem{HawkingPenrose1970}S.W. Hawking and R. Penrose, Proc. Roy. Soc. 
{\bf A246} (1970) 529.

\bibitem{Fotini1998}F. Markopoulou, Commun. Math. Phys. {\bf 211} (2000) 
559; gr-qc/9811053.

\bibitem{GT1990}D.M. Gitman and I.V. Tyutin, {\it Quantization of Fields with
Constraints} (Springer-Verlag, Berlin, 1990).

\bibitem{KKM2}N. Kiriushcheva, S.V. Kuzmin and D.G.C. McKeon, 
Mod. Phys. Lett. {\bf A20} (2005) 1961; hep-th/0503231.

\bibitem{Fotini1996}F.G. Markopoulou, Class. Quant. Grav. {\bf 13} (1996) 
2577. 

\bibitem{Thiemann2003}T. Thiemann, gr-qc/0305080.

\bibitem{Kummer}W. Kummer and H. Sch\"utz, Eur. Phys. J. {\bf C42} (2005) 
227; gr-qc/0410017.

\bibitem{newideas}S.A. Hojman, K. Kuchar, and C.Teitelboim, Ann. Phys. 
{\bf 96} (1976) 88.

\bibitem{Fad1982}L.P. Faddeev, Usp. Fiz. Nauk {\bf 136} (1982) 435; Sov. 
Phys. Usp. {\bf 25} (1982) 130.

\bibitem{FJ}L.P. Faddeev and R. Jackiw, Phys. Rev. Lett. {\bf 60} (1988) 1692. 

\bibitem{Castellani}L. Castellani, Ann. Phys. {\bf 142} (1982) 357.

\bibitem{KM2005}S.V. Kuzmin and D.G.C. McKeon, Ann. Phys. {\bf 318} (2005) 495.

\bibitem{KM2001}S.V. Kuzmin and D.G.C. McKeon, Phys. Rev. {\bf D64} 
(2001) 085009.

\bibitem{Deser}S. Deser, J. McCarthy and Z. Yang, Phys. Lett. {\bf B222} 
(1989) 61.

\bibitem{ADMlast}R. Arnowitt, S. Deser, and C.W. Misner, The dynamics of 
General Relativity, in: {\it Gravitation: An Introduction to Current Research} 
ed. by L. Witten, (Wiley, New York, 1962) 227.

\bibitem{ADM3}R. Arnowitt, S. Deser, and C.W. Misner, Phys. Rev. {\bf 117} 
(1960) 1595.

\bibitem{AshtekarHorowitz1982}A. Ashtekar and G.T. Horowitz, Phys. Rev. 
{\bf D12} (1982) 3342.

\bibitem{DiracVsReduced}J.D. Romano and R.S. Tate, Class. Quant. Grav. {\bf 6} 
(1989) 1487; K. Schleich, Class. Quant. Grav. {\bf 7} (1990) 1529; 
G. Kunstatter, Class. Quant. Grav. {\bf 9} (1992) 1469; R.J. Epp, Phys. Rev. 
{\bf D50} (1994) 6569 and 6578.

\bibitem{DeWitt1967}B.S. DeWitt, Phys. Rev. {\bf 160} (1967) 1113.

\bibitem{Samuel2001}J. Samuel, Phys. Rev. {\bf D63} (2001) 068501.

\bibitem{IshamKuchar}F. Antonsen and F. Markopoulou, gr-qc/9702046.

\bibitem{Woodard}D.A. Eliezer and R.P. Woodard, Nucl. Phys. {\bf B325} (1989) 
389.

\bibitem{newvariables}A. Ashtekar, Phys. Rev. Lett. {\bf 57} (1984) 2244; 
Phys. Rev. {\bf D36} (1987) 1587; J.F. Barbero, Phys. Rev. {\bf D51} (1995) 
5507. 

\bibitem{PlyushchayRazumov}M.S. Plyuschay and A.V. Razumov, Int. J. Mod. Phys.
{\bf A11} (1996) 1427. 

\bibitem{Anderson}E. Anderson, gr-qc/0409123.
\end{thebibliography}
\end{document}